\documentclass{article}
\makeatletter
\def\ps@headings{%
	\def\@oddhead{\mbox{}\scriptsize\rightmark \hfil \thepage}%
	\def\@evenhead{\scriptsize\thepage \hfil \leftmark\mbox{}}%
	\def\@oddfoot{}%
	\def\@evenfoot{}}
\makeatother 

\PassOptionsToPackage{english}{babel}
\usepackage{amsmath,amsfonts,amsthm}
\usepackage{array}
\usepackage[caption=false,font=normalsize,labelfont=sf,textfont=sf]{subfig}
\usepackage{textcomp}
\usepackage{stfloats}
\usepackage{url}
\usepackage{graphicx}
\usepackage{cite}
\usepackage{wrapfig}
\usepackage[table]{xcolor}
\usepackage{booktabs}

\usepackage[urlcolor=blue]{hyperref}
\usepackage{mathtools}
\usepackage{amssymb}
\usepackage{tcolorbox}

\allowdisplaybreaks
\usepackage{tabularx}

\usepackage[font=small,labelfont=bf]{caption}
\usepackage{emptypage}
\usepackage{epigraph}
\usepackage{fancyhdr}
\usepackage{fancyvrb}
\usepackage{float}
\usepackage{fontenc}
\usepackage[left=18mm,
right=18mm,top=20mm,bottom=20mm]{geometry}

\newcommand{\ex}{\textrm{e}}
\newcommand{\expo}[1]{\textsl{exp}\left(#1\right)}
\newcommand{\Hexp}[1]{\textsl{hypoexp}\left(#1\right)}
\newcommand{\inv}[1]{\frac{1}{#1}}

\newcommand{\dd}{\textrm{d}}
\newcommand{\eqdef}{\stackrel{def}{=}}
\newcommand{\E}[1]{\mathbb{E}\left[#1\right]}

\newcommand{\Pb}[1]{\mathbb{P}\left[#1\right]}
\newcommand{\Cov}[1]{\mathbb{C}\mathrm{ov}{}\left[#1\right]}
\newcommand{\cg}{\cellcolor{gray!20}}

\theoremstyle{remark}
\newtheorem*{remark}{Remark}

\theoremstyle{proposition}
\newtheorem{proposition}{Proposition}

\theoremstyle{theorem}
\newtheorem{theorem}{Theorem}

\theoremstyle{lemma}
\newtheorem{lemma}{Lemma}

\theoremstyle{theorem}
\newtheorem*{myProof}{Proof}

\theoremstyle{corollary}
\newtheorem{corollary}{Corollary}

\renewcommand{\qedsymbol}{\qquad_\square}

\title{On the Age of Information in Single-Server Queues with Aged Updates}
\author{Fernando Miguelez$^a$$^b$ \and Urtzi Ayesta$^{c,d,e,f}$ \and Josu Doncel$^g$ \and Maria Dolores Ugarte$^a$$^b$}

\date {%
	\small
	$^a$ Department of Statistics, Computer Science and Mathematics, Public University of Navarre\\%
	$^b$ Institute for Advanced Materials and Mathematics (InaMat$^2$)\\%
	$^c$ Department of Computer Science, University of the Basque Country, EHU\\
	$^d$ IkerBasque - Basque Foundation for Science
	$^e$ CNRS, IRIT
	$^f$ Institut National Polytechnique, Toulouse\\%
	$^g$ Department of Mathematics, University of the Basque Country, EHU\\%
}

\begin{document}

\maketitle

\begin{abstract}
	
The Age of Information (AoI) is a performance metric that quantifies the freshness of data in systems where timely updates are critical. Most state-of-the-art methods typically assume that packets enter the monitored system with zero age, neglecting situations, such as those prevalent in multi-hop networks or distributed sensing, where packets experience prior delays. In this paper, the AoI is investigated when packets have a non-zero initial age. We derive an expression for the average AoI in this setting, showing that it equals the standard AoI plus a correction term involving the correlation between packet age and inter-departure times. When these variables are independent, the expression simplifies to an additive correction equal to the mean initial age. In cases where the dependency structure is unknown, we also establish lower and upper bounds for the correction term. We demonstrate the applicability of our approach across various queueing scenarios. We recover known results for systems like forwarding and homogeneous tandem queues, and derive the first AoI expressions for two novel models: the heterogeneous M/M/1/1 $\rightarrow$ $\cdot$/M/1/$\infty$ tandem and the M/M/1 retrial queue. Additionally, we explore the accuracy of the derived bounds on a tandem composed of several queues, a model that has not yet been analytically solved from an age perspective. 
	
\end{abstract}

\section{Introduction}
The Age of Information (AoI) is a relatively new metric that measures the freshness of the knowledge we have about the status of a remote system. More specifically, the AoI of a process is defined as the time elapsed since the generation of the last successfully received packet containing information about that process. The emergence of AoI in the seminal paper \cite{kaul:2011} has triggered an interest in its analysis in the context of queueing theory, notably motivated by the fact that policies that optimise classical performance metrics, such as throughput, delay or package-loss probability, do not necessarily minimise the AoI.

\begin{figure}[htpb]
	\centering
	\begin{minipage}{0.45\textwidth}
		\centering
		\includegraphics[width=\textwidth, clip=true, trim=0cm 14.6cm 17cm 1.5cm]{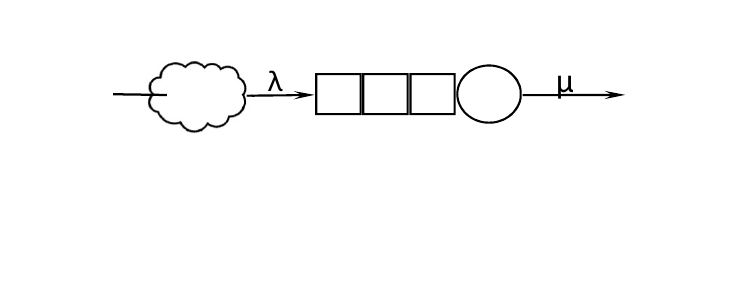}
		
		\vspace{-.8cm}
		\footnotesize{(a)}
	\end{minipage}%
	\begin{minipage}{0.45\textwidth}
		\centering
		\includegraphics[width=\textwidth, clip=true, trim=1.4cm 14.83cm 15.2cm 2.17cm]{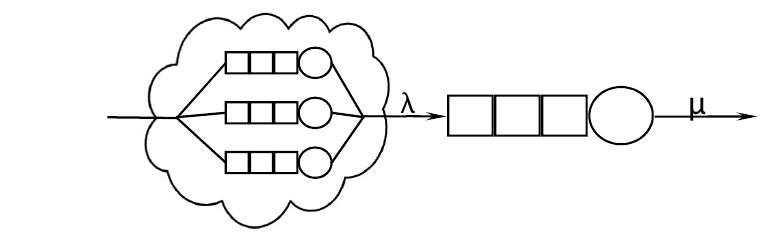}
		\footnotesize{(b.1)}
	\end{minipage}
	
	\vspace{0.5cm} 
	
	\begin{minipage}{0.45\textwidth}
		\centering
		\includegraphics[width=\textwidth, clip=true, trim=2.11cm 15.46cm 14.86cm 1.8cm]{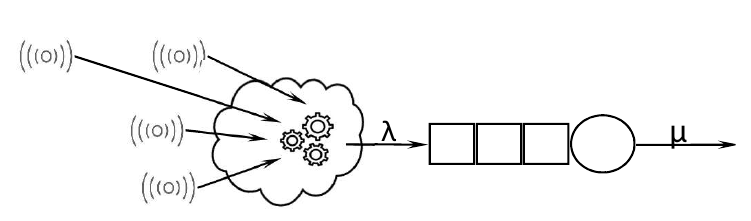}
		
		\vspace{0.5cm}
		\footnotesize{(b.2)}
	\end{minipage}
	\begin{minipage}{0.45\textwidth}
		\centering
		\includegraphics[width=\textwidth, clip=true, trim=1.5cm 16.02cm 15.65cm 1.2cm]{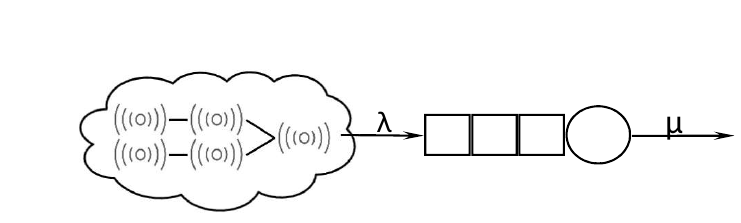}
		
		\vspace{0.5cm}
		\footnotesize{(b.3)}
	\end{minipage}
	
	\caption{Motivating example for the proposed aged updates framework with some application examples.}
	\label{fig::motivation_example}
\end{figure}

Most of the research in AoI considers the case in which the age of the packets to be delivered is zero, which corresponds to packets entering the communication channel as soon as they are generated. Under this assumption, a rich body of literature has developed, providing a variety of analytical tools and foundational results. This includes the foundational analysis of the average AoI in various queueing systems using methods such as the graphical approach \cite{kaul_real:2012, costa:2016, yates:2019, soysal:2021, lin:2022} and Stochastic Hybrid Systems (SHS) \cite{yates:2018, yates:2019, yates:2020, kam:2022}, optimal scheduling policies aimed at minimising AoI \cite{bedewy:2019, yates:2019}, and investigations into AoI behaviour under energy harvesting constraints \cite{arafa:2019}. The field has also expanded beyond just the average AoI to explore a family of age-related metrics, such as the peak AoI \cite{costa:2016}, the Age of Incorrect Information \cite{maatouk:2020}, and the Query AoI \cite{chiariotti:2022}. Additionally, researchers have examined these concepts through game-theoretic perspectives \cite{nguyen:2018, miguelez:2023}. More recent and general analytical efforts have focused on deriving the full probability distribution of the AoI \cite{inoue:2019, yates:2020} and establishing performance bounds for a wider range of systems \cite{farazi:2019, soysal:2021}. Comprehensive surveys \cite{sun:2020, yates:2021} and texts \cite{kosta:2017} document the rapid expansion of this field. A repository of papers related to AoI can be found at \cite{AoIRepository}.

In this paper, the focus is on cases where packets to be delivered have a non-zero age. This situation arises when packets have already spent time in another system before reaching the final link, which is common in multi-hop networks, distributed sensing, and edge computing. For instance, in a wireless network, updates may be generated at a sensor, relayed through multiple intermediate nodes, and eventually delivered to a destination. By the time they enter a specific queue, they have already accumulated some age from prior hops. Another example is found in a sensor network, in which sensors collect and locally process data before transmitting it to a remote server. In such cases, the AoI at the receiver reflects both the local processing time and the transmission delays. \autoref{fig::motivation_example}(a) depicts an elementary example of this model with some potential application scenarios. Between the packet's generation time and its arrival at the system queue, some delay exists, which may be due to various causes, such as intermediate buffers (b.1), network latency or preprocessing (b.2), or sensor remoteness in a multi-hop network (b.3). This delay produces a sequence of initial ages on the incoming packets.

The analysis of AoI in multi-stage systems is gaining significant attention, with recent works making substantial progress in specific directions. These include the quantitative characterisation of AAoI in tandems of queues \cite{kam:2022}, developing scheduling policies for multi-hop wireless networks \cite{talak:2017}, characterising age accumulation under preemptive service disciplines \cite{yates:2018}, deriving bounds for systems with cross-traffic \cite{vikhrova:2022}, and exploring advanced concepts like optimisation under imperfect knowledge \cite{zhao:2025}, peak-age analysis in cache-enabled networks \cite{ernest:2025}, and semantics-aware status updates \cite{delfani:2025}.

However, a fundamental challenge remains: the conventional approach of modelling each multi-stage system from scratch often leads to highly specific and potentially intractable derivations. As a result, there is a scarcity of general analytical results that can be applied across different system configurations, as acknowledged in \cite{farazi:2019}. The proposal presented in this paper can therefore be seen as a novel view to address this challenge. Instead of analysing the entire system complexity at once, we model the final communication link as a single-server queue where packets arrive with a non-zero initial age, effectively decoupling the analysis of the prior system's impact from the dynamics of the final queue.

In our main contribution, we provide an expression for the Average Age of Information (AAoI) when packets have a non-zero initial age ($\Delta^A$), relating it directly to the AAoI in the standard case where packets arrive with zero age ($\Delta^0$). The resulting decomposition, given in Theorem~\ref{thm::aaoi_au}, serves as a unified tool to tackle a wide class of problems, enabling both the recovery of known results and the derivation of new expressions and bounds for complex systems where traditional methods struggle.

It is important to note that the proposed framework is designed to be general and does not assume any specific packet management policy, such as discarding obsolete packets at the monitor. This choice is deliberate for two reasons. First, it allows the model to be applied to systems where such discarding is not feasible, such as those requiring in-order delivery for state consistency, complete data reconstruction, or non-volatile logging. Second, and more importantly, the effect of any packet management policy is inherently captured by our model through its impact on the effective arrival rate, $\lambda^e$, and the dependence structure between the initial ages and the departure process at the final queue. This versatility is a key strength of the proposed approach.

The main contributions of this work are:
\begin{itemize}
	\item \textbf{General framework:} We derive an expression for the AAoI in a single-server queue with aged updates (\autoref{thm::aaoi_au}). The result shows that the AAoI equals the standard AAoI plus a correction term involving the correlation between the initial age and inter-departure times. When these variables are independent, the expression simplifies to an additive correction equal to the mean initial age.
	\item \textbf{Performance bounds:} For cases where the dependency structure is unknown, we establish lower and upper bounds for the correction term (Corollary~\ref{cor::interval}) that depend only on the marginal distributions of the involved processes.
	\item \textbf{Unification of results:} We demonstrate that the framework can recover AAoI expressions for well-studied systems, such as forwarding queues and tandem queues, under a common analytical approach.
	\item \textbf{Novel characterisations:} We provide the first AoI analysis for two specific models: The heterogeneous M/M/1/1 $\rightarrow$ $\cdot$/M/1/$\infty$ tandem queue, and the M/M/1 retrial queue with single retrials.
	\item \textbf{Practical utility:} We explore the accuracy of the derived bounds in complex scenarios, such as a tandem of multiple queues, and validate their effectiveness through simulations.
\end{itemize}

The rest of the paper is organised as follows. In \autoref{sec::model}, the system model is introduced and the age process with aged updates is characterised. In \autoref{sec::result}, the main result of this article is presented and demonstrated. In \autoref{sec::applications}, some illustrative applications of the model are described. Finally, some final comments are given in \autoref{sec::conclusions} together with some suggestions for future research.
\section{Model Description}\label{sec::model}
\subsection{System model}
Consider a communication system through which a destination node receives updates about a process of interest. Information updates are generated at the source at epochs $g_n$ and transmitted to the destination as packets. Unlike standard models, these updates do not enter the queueing system immediately upon generation but arrive after incurring a certain delay.

Packets enter the queueing system at epochs $t_n$ according to an arrival process of rate $\lambda$ with i.i.d. inter-arrival times $X_n = t_n - t_{n-1}$. Each arriving packet has a corresponding initial age $A_n \geq 0$, defined as the time elapsed between the packet’s generation and its arrival at the system. The sequence of initial ages $\{A_n\}$ is assumed to be stationary and ergodic, with identically distributed (but not necessarily independent) random variables. Importantly, $A_n$ can be statistically dependent on the inter-arrival times $X_n$ and the future inter-departure times $Y_n$. This dependence is a key aspect of our framework, reflecting realistic scenarios and introducing additional analytical complexity. The stationarity assumption ensures the process remains well-behaved, avoiding cases of ever-increasing ages.

Upon arrival, a packet is either discarded, queued for future processing, or served immediately, depending on the queue size, the packet management policy, and the system's state. Packets that are not discarded enter the system and complete the service. These are referred to as \emph{effective arrivals}, and the effective inter-arrival times are denoted by $X_n^e$. Once a packet enters the server, it stays in service for a time $S_n\sim\expo{\mu}$, independent of the arrival process. The service completion times are denoted by $t'_n$, and define the sequence of departures. The \emph{system time} of packet $n$ is $T_n=t'_n-t_n$, and the \emph{inter-departure time} is $Y_n=t'_n-t'_{n-1}$. 

The timeliness of information at the destination node can be quantified using the metric AoI, which is discussed in the next section.
\subsection{Age of Information with aged updates}

Define the counting process of delivered packets 
\begin{equation}\label{def::counting_process}
	N(s)=\max\left\{n: t'_n\leq s\right\}.
\end{equation} 

At any given time $s$, the timestamp of the last packet received at the destination is $g_{N(s)}$. The AoI is thus defined as the random process
\begin{equation}\label{def::age_def}
	\Delta(s) = s - g_{N(s)},\ s>0.
\end{equation}

\begin{figure}[htpb]
	\centering
	\includegraphics[width=0.58\textwidth]{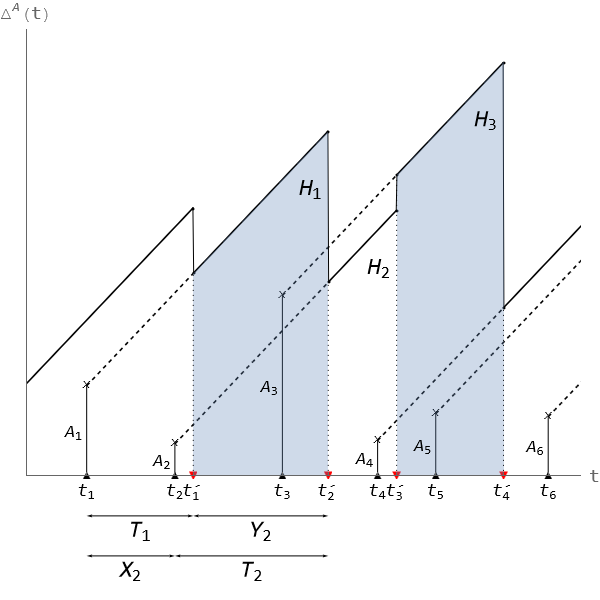}
	\caption{Example of age profile in a model with aged updates. The packet arriving at epoch $t_3$ carries a high age. When the packet is delivered at $t'_3$, it produces an upper jump on the age at the monitor. Disjoint areas $H_n$ are used to calculate the Average Age of Information.}\label{fig::aoi_path}
\end{figure}

A graphical illustration of the AoI evolution in this model is shown in \autoref{fig::aoi_path}. The process evolves piecewise linearly, with occasional jumps at delivery times $t'_n$. While it still maintains the typical sawtooth shape of age functions, some updates may enter the system with a high initial age, meaning they carry obsolete information. At the time of delivery of these updates, the age at the destination may increase instead of resetting to a lower value. This behaviour is a distinctive feature of models with aged updates and is not seen in standard AoI models with zero-aged updates \cite{yates:2021}.
Specifically, at times $t'_n$, the AoI resets from the peak value $A_{n-1}+X_n+T_n$ to $A_n+T_n$. Thus, if $A_n$ exceeds the threshold $X_n+A_{n-1}$, the delivery of this packet leads to an upper jump in the AoI function, as illustrated by packet 3 in \autoref{fig::aoi_path}. We term these ''far updates'' due to their similarity to distant status updates in asynchronous monitoring. 

The occurrence of far updates, while detrimental to age performance, is unavoidable in systems where packets cannot be discarded, such as those requiring in-order delivery (e.g., distributed consensus), data reconstruction (e.g., media streaming with error correction), or maintaining complete historical records (e.g., flight data recorders). Although average freshness may not be the primary concern in some of these applications, tracking the behaviour of the AoI can reveal latent flaws or performance bottlenecks in the system. The model is also relevant when AoI itself is the primary performance metric of the system, regardless of the information contained in the packet. In this context, discarding a high-age packet would artificially improve the metric and mask the true system performance. Furthermore, many systems naturally prevent the existence of far updates through their inherent dynamics, as demonstrated in Sections IV-C.1 and IV-C.3.

On the other hand, when packet discarding is implemented, its effect is inherently captured by our framework through the effective arrival rate $\lambda^e$ and the dependency structure between $A_n$ and $Y_n$. Thus, the proposed approach remains relevant regardless of the packet management policy.
\subsection{Notation}
\autoref{tab::notation} contains the main notation used throughout this article.
\begin{table}[h]
	\centering
	\caption{Summary of notation used in the model description}
	\begin{tabular}{ll}
		\toprule
		\textbf{Symbol} & \textbf{Description} \\
		\midrule
		$A_n$         & initial age of the $n$-th packet upon arrival \\
		$t_n$         & arrival time of the $n$-th packet \\
		$t_n'$        & service completion (departure) time of the $n$-th packet \\
		$t_n''$       & service completion of the $n$-th packet at the second queue (tandem queues)\\
		$t^o_n$       & arrival at orbit of the $n$-th packet (retrial queues)\\
		$g_n$         & generation time of the $n$-th packet, $g_n = t_n-A_n$ \\
		$X_n$         & inter-arrival time between real arrivals, $X_n = t_n - t_{n-1}$ \\
		$X_n^e$       & inter-arrival time between effective arrivals (admitted into service) \\
		$S_n$         & service time of the $n$-th packet, $S_n\sim\expo{\mu}$ \\
		$T_n$         & system time of the $n$-th packet, $T_n = t_n'-t_n$ \\
		$Y_n$         & inter-departure time, $Y_n = t_n'-t_{n-1}'$ \\
		$N(s)$        & index of the last packet delivered by time $s$ \\
		$\Delta^A(s)$ & Age of Information at time $s$, $\Delta^A(s) = s-g_{N(s)}$ with aged updates\\
		$\Delta^A$    & Average Age of Information with aged updates\\
		$\Delta^0$    & Average Age of Information with zero-aged updates\\
		$d$           & interval width (Corollary \ref{cor::interval})\\
		$\lambda$     & arrival rate of packets \\
		$\lambda^e$   & effective arrival rate (of admitted packets) \\
		$\mu$         & service rate (queue of interest)\\
		$\gamma,\gamma_i$ & service rates (prior queues in tandems)\\
		$\theta$      & retrial rate (retrial queues)\\
		\bottomrule
	\end{tabular}
	\label{tab::notation}
\end{table}
\section{Main results}\label{sec::result}
The following theorem states the main result of this article.

\begin{theorem}\label{thm::aaoi_au}
	For the single-server queue with aged updates described in Section \ref{sec::model}, the AAoI is given by
	\begin{equation*}
		\Delta^A=\Delta^0+\lambda^e\E{Y_n A_{n-1}},
	\end{equation*}
	where $\Delta^0$ is the AAoI in the corresponding system with zero-aged updates, $\lambda^e\eqdef1/\E{X^e}$ is the effective arrival rate (i.e., the rate of packets that are admitted into service), $Y_n$ is the $n$-th inter-departure time, and $A_{n-1}$ is the initial age of the $(n-1)$-th packet.
\end{theorem}

\begin{myProof}
	The graphical method is adopted, a widely used technique for computing the AAoI via geometric area decomposition \cite{costa:2016,yates:2019,yates:2021}. Let us examine \autoref{fig::aoi_path}, which illustrates the evolution of the AoI in a system with aged updates, including a sample far update, and suppose that the process is observed during an interval $[0,T]$. Without loss of generality, it can be assumed that deliveries occur at both the initial time $t=0$ and the final time $t=T$, that is, $t'_0=0$ and $t'_n=T$ for some $n\leq N(T)$, where $N(T)$ is the number of delivered updates at the end of the observed period, as defined in \eqref{def::counting_process}. This is a harmless assumption since the only consequence of leaving it out is the addition of two finite areas that, nonetheless, will vanish in the limit. Also, for simplicity, assume that after the delivery at $t=0$, the system becomes empty. 
	
	Under the ergodicity assumption, the average of the process $\Delta^A(t)$ along the observed interval can be calculated as the time average
	\begin{equation}\label{eq::aaoi_T1}
		\Delta_T^A=\dfrac{1}{T}\int_0^T\Delta^A(t)\dd t
	\end{equation}
	
	However, the integral in \autoref{eq::aaoi_T1} can be replaced by the sum of the disjoint polygonal areas $H_n$, for $n=0,1,\ldots,N(T)-1$. Each of these areas consists of the sum of a triangular area and a rectangular area,
	\begin{equation}
		H_n=\dfrac{Y_{n+1}^2}{2} + Y_{n+1}(A_n+T_n).
	\end{equation} 
	
	Hence, \autoref{eq::aaoi_T1} can be written as
	\begin{equation}\label{eq::aaoi_T2}
		\Delta_T^A=\dfrac{1}{T}\sum\limits_{n=0}^{N(T)-1}H_n=\dfrac{1}{T}\sum\limits_{n=0}^{N(T)-1}\left(\dfrac{Y_{n+1}^2}{2} + Y_{n+1}(A_n+T_n)\right).
	\end{equation}
	
	The AAoI is then computed taking \eqref{eq::aaoi_T2} to the limit as $T\to\infty$, 
	\begin{equation}\label{eq::aaoi_limit}
		\Delta^A=\lim_{T\to\infty}\Delta_T^A.
	\end{equation} 
	
	Under the stability condition $\E{S_n} < \E{X_n^e}$, the system reaches a steady state, and the sequence of effective inter-arrival times $\{X_n^e\}$ forms a stationary and ergodic process. By the principle of flow conservation \cite{krakowski:1973}, the long-term effective arrival rate equals the long-term departure rate. Additionally, the Elementary Renewal Theorem \cite{ross:stochastic} asserts that the ratio between the number of delivered packets and the interval length approaches the departure rate. Therefore, $$\lambda^e=1/\E{X_n^e}=1/\E{Y_n}=\lim_{T\to\infty}\dfrac{N(T)}{T}.$$
	
	The limit in \autoref{eq::aaoi_limit} can then be expanded as
	\begin{align*} 
		\Delta^A &= \lim_{T\to\infty}\dfrac{1}{T}\sum\limits_{n=0}^{N(T)-1}\left(\dfrac{Y_{n+1}^2}{2} + Y_{n+1}(A_n+T_n)\right)\\&\\
		&= \lim_{T\to\infty}\dfrac{N(T)}{T}\ \lim_{T\to\infty}\dfrac{1}{N(T)}\sum\limits_{n=0}^{N(T)-1}\left(\dfrac{Y_{n+1}^2}{2} + Y_{n+1}(A_n+T_n)\right)\\&\\
		&= \lambda^e\left(\E{Y_{n+1}^2}/2+\E{Y_{n+1}T_n}+\E{Y_{n+1}A_n}\right).
	\end{align*}
	
	The first two terms in the parentheses are recognised as those defining the standard AAoI with zero-age updates \cite{yates:2021, costa:2016}, and the proof is complete $\qedsymbol$.
\end{myProof}

\begin{remark}\label{rem:discard_policy}
	The framework naturally incorporates the effect of packet management policies, such as blocking or discarding updates at the monitor. The impact of any such policy is directly captured by two elements in Theorem 1:
	\begin{enumerate}
		\item[a)] The \emph{effective arrival rate} $\lambda^e$, which reflects only the packets admitted into the system. A discard policy would reduce $\lambda^e$ compared to the raw arrival rate $\lambda$.
		\item[b)] The \emph{dependence structure} between $Y_n$ and $A_{n-1}$, as the policy determines which packets are served and thus defines the joint distribution of inter-departure times and initial ages.
	\end{enumerate}
	Therefore, our general expression is also applicable in the case where any of these policies exists, providing a foundation for their analysis rather than being incompatible with them.
\end{remark}

A well-known identity in probability says that the expected value of the product of two random variables can be expressed as the sum of their covariance and the product of their means: $\E{Y_nA_{n-1}}=\Cov{Y_n,A_{n-1}}+\E{Y_n}\E{A_n}$. Moreover, the coefficient of variation of a random variable is defined as the ratio between the standard deviation and the mean, $\kappa_Y=\sigma_Y/\E{Y_n}$, while the correlation coefficient between two random variables is the ratio between their covariance and the product of the standard deviations, $r_{Y,A}=\Cov{Y_n,A_{n-1}}/(\sigma_Y\sigma_A)$, where $\sigma_\cdot$ stands for the standard deviation of the corresponding random variable. Because of this, the age correction term in \autoref{thm::aaoi_au} can be rewritten as

\begin{align}
	\lambda^e\E{Y_nA_{n-1}} &= \dfrac{\Cov{Y_n,A_{n-1}}+\E{A_n}\E{Y_n}}{\E{Y_n}}\nonumber\\
	&= \E{A_n} + \dfrac{r_{Y,A}\cdot\sigma_Y\cdot\sigma_A}{\E{Y_n}}\nonumber\\
	&= \E{A_n} + r_{Y,A}\cdot\kappa_Y\cdot\sigma_A\label{eq::corr_term2}.
\end{align} 

\begin{remark}
	The above expression states that, given the marginal distributions of $Y_n$ and $A_{n-1}$, the dependence between them affects the correction term only through the correlation coefficient. From another perspective, \autoref{eq::corr_term2} can also be interpreted as implicitly limiting the correlation between $Y_n$ and $A_{n-1}$. To understand this, we first note that the correction term is always non-negative, because it is the product of a positive rate and the expected value of a non-negative random variable, $Y_nA_{n-1}$. Intuitively, the non-zero initial age $A_{n-1}$ represents a pre-existing delay that can only contribute to an increase in the overall age at the monitor, as it represents time that has already elapsed and cannot be ``recovered''. The term $\E{Y_nA_{n-1}}$ quantifies the interaction between this prior delay and the system's dynamics, always adding to the baseline age $\Delta^0$.  Therefore, $\E{A_n} + r_{Y,A}\cdot\kappa_Y\cdot\sigma_A>0$, and the correlation coefficient must satisfy $r_{Y,A}>-\kappa_A^{-1}\kappa_Y^{-1}$.
	
	This inequality implies specific tendencies for the correlation: it tends to be positive if either variable shows high variability or negligible mean (i.e., $\kappa_Y\to \infty$ or $\kappa_A\to \infty$). Conversely, a strong negative correlation is only possible when, for given means, at least one of them is highly stable ($\sigma_A\to 0$ or $\sigma_Y\to 0$), or, for given standard deviations, at least one of them exhibits very large average values ($\mu_A\to\infty$ or $\mu_Y\to\infty$).
\end{remark}

The following corollary is also a direct consequence of \autoref{eq::corr_term2}.

\begin{corollary}\label{cor::interval}
	In any queueing system with inter-departure times $Y_n$ and initial ages $A_n$, the correction term in \autoref{thm::aaoi_au} is bounded by
	\begin{equation*}
		\E{A_n} - d/2 \leq \lambda^e\E{Y_nA_{n-1}} \leq \E{A_n} + d/2,
	\end{equation*} 
	where $d=2\kappa_Y\sigma_A$ is the interval width.
\end{corollary}

The lower and upper bounds provided above depend exclusively on the marginals of $Y_n$ and $A_n$, which makes them particularly helpful for anticipating the correction term's value in scenarios where the dependence structure between $Y_n$ and $A_{n-1}$ is unknown or not analytically tractable.
\section{Applications}\label{sec::applications}
In this section, \autoref{thm::aaoi_au} is applied to several queueing scenarios to motivate the versatility and relevance of the aged update framework. Initially, the simple case is examined where each update experiences a fixed delay before entering the system. Next, a system where retransmissions occur after transmission errors is considered. Here, the delay from repeated attempts implicitly induces a process of initial ages. The computation of the AAoI at the output of tandem queueing systems is also analysed, where the time spent in previous nodes naturally contributes to the age at the final queue. Both a classical case involving identical queues and a more general setting with heterogeneous queues are addressed, highlighting the flexibility of the proposed approach across different network topologies. Finally, an example of a retrial queue is considered, in which the time spent in orbit is interpreted as the initial age for updates in an analogous model. When deemed interesting, the tightness of the bounds provided in \autoref{cor::interval} will be assessed analytically and numerically.
\subsection{Independency}\label{sec::independency}
We first examine the relevant special case in which the sequence ${A_n}$ of initial ages is independent of the sequence of inter-departure times ${Y_n}$. Under this assumption, the model analysis greatly simplifies since the cross term of \autoref{thm::aaoi_au} factorises as $\E{Y_nA_{n-1}}=\E{Y_n}\E{A_{n-1}}$, and the AAoI can then be expressed as

\begin{equation*}
	\Delta^A=\Delta^0+\E{A_n}.
\end{equation*}

An elementary and illustrative example occurs when each arriving packet undergoes a fixed delay $A>0$ before entering the system. In this setting, all updates have the same initial age $A_n=A$, leading to $$\Delta^A=\Delta^0 + A.$$ This case is relevant in real-world applications where sources and monitors are not perfectly synchronised. For instance, \cite{beytur:2019} reports observations of fixed delays in practical communication systems due to clock shifts. Our framework naturally incorporates such delays, allowing for a direct and simple quantification of their impact on information freshness.

The potential of the independent case also emerges when the correction term is examined in the single-server FCFS queue under traffic's limiting conditions. In particular, its behaviour in the \emph{heavy traffic} (HT) regime, in which the server is nearly saturated, and the \emph{light traffic} (LT) regime, in which the server utilisation rate is close to zero, allows for the anticipation of the effect of aged updates on freshness in stressed or idle environments. An approximation of the value of the correction term under these two regimes is discussed next.
\subsubsection{Heavy traffic}
In the HT regime, the single-server FCFS queue operates nearly continuously, with negligible idle periods between jobs. Moreover, the Lindley-type equation
\begin{equation}\label{eq::Yn_identity}
	Y_n=S_n+(X^e_n-T_{n-1})^+,
\end{equation}
where $x^+=\max\{0,x\}$, defines the inter-departure times as the sum of the service and the idle times. As a result, under HT conditions, the inter-departure times asymptotically converge in distribution to the service time distribution, and therefore become independent of the arrival process and initial ages. Consequently, in \autoref{eq::corr_term2}, $r_{Y,A}\to0$, and the correction term collapses to $\E{A_n}$. Furthermore, since $\kappa_Y\to 1$, the range of the interval defined in \autoref{cor::interval} approaches $d\to2\sigma_A$, resulting in a tighter interval for stable age processes.
\subsubsection{Light traffic}
The LT regime occurs when the effective arrival rate approaches zero. Together with the flow conservation principle, this feature is captured by the condition $$\E{Y_n}=\E{X_n^e}\to\infty.$$ For a given standard deviation, this implies that the coefficient of variation $\kappa_Y$ goes to zero. Consequently, the second term in the right-hand side of \autoref{eq::corr_term2} and the interval width in \autoref{cor::interval} become negligible, which leads to the independent case with very tight bounds.
\subsection{Forwarding}
The computation of AAoI in error-prone models, where there is a forwarding protocol after a failed transmission, can be performed using the scheme of aged updates. Intuitively, a failed transmission has a dual effect on the age process. On one hand, a failure does not improve the age process because the information has not been received correctly. On the other hand, a forwarded packet has already spent some time in the system, which makes it indistinguishable, from the perspective of information timeliness, from a packet that arrives with an initial age.

To illustrate this, a zero-wait model is considered, in which the source is aware of the system's state and can generate a fresh packet as soon as a delivery occurs. This model is idealised, but also insightful in capturing the best-case AoI performance \cite[Section III.B]{yates:2021}. In particular, because of the source responsiveness, the time variables in the zero-wait model are identically distributed,

\begin{equation}\label{eq::equal_dist_zw}
	X_{n+1}=T_n=Y_n=S_n.
\end{equation}

\begin{figure}[htpb]\captionsetup[subfigure]{font=footnotesize}
	\centering
	\subfloat[]{\includegraphics[width=0.48\textwidth]{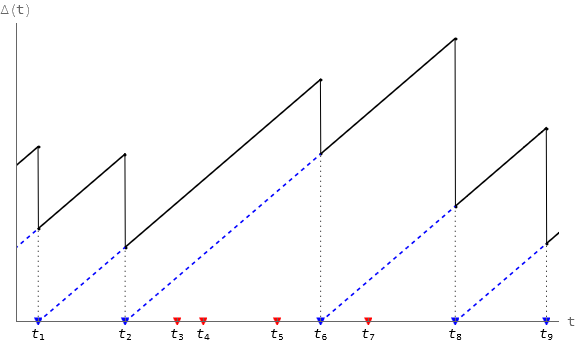}}\qquad
	\subfloat[]{\includegraphics[width=0.48\textwidth]{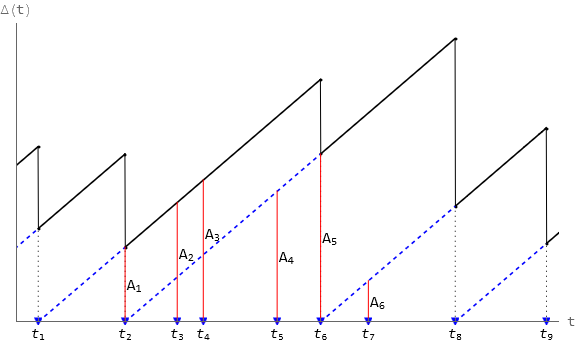}}
	\caption{Error-prone, zero-wait model as a system with aged updates. a) Error-prone, zero-wait standard model. Blue triangles are successful deliveries and generation of a fresh packet, and red triangles ($t_3$ and $t_5$) mark failed deliveries and subsequent retransmission of the same packet. b) Equivalent error-free, zero-wait model with aged updates. Failed transmissions in the standard model can be thought of as inducing an independent process of initial ages $A_n$.}\label{fig::zw_model}
\end{figure}

Suppose further that the communication link is not completely reliable, and a packet arrives corrupted at the destination with probability $\alpha$. The monitor provides acknowledgement (ACK) or negative acknowledgement (NACK) feedback after each update, which is received by the source instantaneously. Upon successful transmission, the source sends a fresh packet; otherwise, it retransmits the same packet --in particular, with the same timestamp-- until an ACK is received. \autoref{fig::zw_model}a) shows how the age evolves in this model. Successful deliveries are marked with a blue triangle and produce a downward jump in the age function. Instead, deliveries marked with a red triangle do not affect the age process, as these deliveries are not completed. \autoref{fig::zw_model}b) presents an equivalent error-free aged-update model, where failed transmissions in the error-prone model can be viewed as triggering an independent initial age process $A_n$.

The value of the AAoI in a zero-wait model with error probability $\alpha$, as that depicted in \autoref{fig::zw_model}a), is given in the following Proposition.
\begin{proposition}\label{prop::zw_aaoi}
	The AAoI in a zero-wait, error-prone model with error probability $\alpha$ and exponential service times of rate $\mu$ is 
	\begin{equation}
		\Delta_{zw}=\dfrac{2}{\mu(1-\alpha)}.
	\end{equation}
\end{proposition}

Note that the number of retransmissions for a given packet in this model is a geometric random variable with parameter $\alpha$. Therefore, the \emph{effective} service time is a geometric sum of exponential times, which results in an exponential distribution of rate $\mu(1-\alpha)$. The result in \autoref{prop::zw_aaoi} then follows directly from \cite[Equation (22)]{yates:2021}.

Notably, the same expression is obtained by applying \autoref{thm::aaoi_au} to the error-free model of \autoref{fig::zw_model}b) with an independent process of initial ages defined by

\begin{equation*}
	A_n=\begin{cases}0&\text{with probability }(1-\alpha)^2\\
		\tilde S_1&\text{with probability }2\alpha(1-\alpha)\\
		\tilde S_2&\text{with probability }\alpha^2\end{cases},
\end{equation*}
where $\tilde S_1\sim\expo{\mu(1-\alpha)}$ and $\tilde S_2\sim\Gamma\left(2,\mu(1-\alpha)\right)$ are implicitly derived from the number of retransmissions. In Appendix \ref{app::appendix3} we provide full details of the proof of \autoref{prop::zw_aaoi} using the aged-updates framework.

\begin{wrapfigure}{r}{0.5\textwidth}
	\centering
	\includegraphics[width=0.45\textwidth]{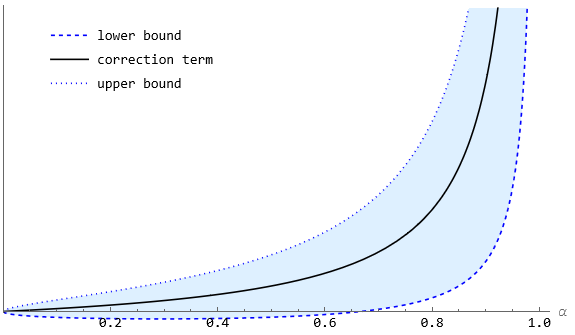}	
	\caption{\footnotesize{Correction term and bounds provided by \autoref{cor::interval} in the error-prone, zero-wait model.}}\label{fig::ct_bounds_zw}
\end{wrapfigure}

While the correction term of this model can be obtained explicitly, the bounds defined in \autoref{cor::interval} remain valuable to gain insight into their tightness. In particular, with a few calculations, it can be found that the width of the interval is given by $$d=2\sigma_A=\dfrac{2\sqrt{2\alpha(2-\alpha)}}{\mu(1-\alpha)}.$$ However, a more detailed understanding is achieved by examining the bounds directly. To this purpose, \autoref{fig::ct_bounds_zw} plots the correction term and the upper and lower bounds against the error probability $\alpha$, for a fixed service rate $\mu=1$.

The plot reveals some key insights. The bounds are generally tight for low error probabilities, where successful transmissions dominate the system behaviour. As $\alpha$ increases, the upper bound grows considerably, reflecting the high uncertainty and variability introduced by frequent retransmissions. Furthermore, the lower bound becomes negative for $\alpha<2/3$. In this region, the trivial bound $\lambda^e\E{Y_nA_{n-1}}\geq0$ is better for practical purposes. This occurs because the high variability of the initial age $\sigma_A$ outweighs its mean value for moderate error rates.
\subsection{Tandem Queues}\label{sec:tandems}
Aged updates also prove useful for computing the AAoI in general queueing networks. In such configurations, the age of a packet upon arrival reflects the time spent at earlier nodes. A particularly illustrative case is that of a tandem queueing system, where the final AAoI can be obtained by separately evaluating the AAoI at the last queue and then adding a correction term, as indicated by \autoref{thm::aaoi_au}, for the correlation between the final inter-departure times and the initial ages.

Some prior works have derived bounds for age metrics in multi-hop systems. For instance, \cite{farazi:2019} establishes fundamental lower bounds on the peak and average age for general multi-source, multi-hop wireless networks under different interference constraints. Similarly, \cite{vikhrova:2022} derives bounds on the average AoI for systems with cross-traffic and various queuing policies. In this section, we demonstrate that our methodology, which relies on a correlation-based decomposition at a single queue, provides a complementary and effective tool for analysing tandem networks.

The following examples illustrate the usefulness of this idea. In the first example, a tandem of two infinite-capacity M/M/1 is studied. While the final AAoI in this setup is already known, it validates the application of the aged-update framework in the presence of statistical dependence between the inter-departure times and the initial ages. In the next example, the previous tandem is extended to an arbitrary number of queues, to highlight the practical advantages of \autoref{cor::interval}. The last example addresses a tandem formed by two heterogeneous queues, a system barely explored in the literature compared to those composed of identical queues. 

Throughout these examples, superscripts $(1)$ and $(2)$ are included when needed to identify the variables in each tandem queue. 
\subsubsection{M/M/1/$\infty$ $\rightarrow$ M/M/1/$\infty$ tandem}\label{sec::tandem2}
This first tandem model serves as a primary illustration of the framework's ability to handle statistical dependence between the initial ages, which here represents the time spent in the first queue of the tandem, and the inter-departure times in the second queue. This dependence is captured by the explicit computation of the cross-term $\E{Y_n^{(2}A_{n-1}}$ in \autoref{lemma::ET1Y2}. First, the theorem that gives the AAoI at the end of this tandem is stated. While this result is already known, this example is useful for confirming that the approach with aged updates can also be accommodated in this model.

\begin{theorem}[\cite{kam:2022}, Theorem 3\footnote{\cite[Equation (13)]{kam:2022} contains a typo.}]\label{thm::aaoi_tandem0}
	The AAoI at the end of two infinite-capacity queues in tandem (M/M/1/$\infty$ $\rightarrow$ M/M/1/$\infty$) with $\lambda$-Poisson arrivals, and i.i.d. exponential service times of rates $\gamma$ and $\mu$, is given by
	
	\begin{equation*}
		\Delta_{M/M/1/\infty \rightarrow M/M/1/\infty} = \inv{\lambda} + \inv{\mu} + \dfrac{\lambda^2/\mu^2}{\mu-\lambda} + \inv{\gamma} + \dfrac{\lambda^2/\gamma^2}{\gamma-\lambda} + \dfrac{\lambda^2/\gamma\mu}{\gamma+\mu-\lambda}.
	\end{equation*}
\end{theorem}

\begin{figure}[htpb]
	\centering
	\includegraphics[width=0.75\textwidth]{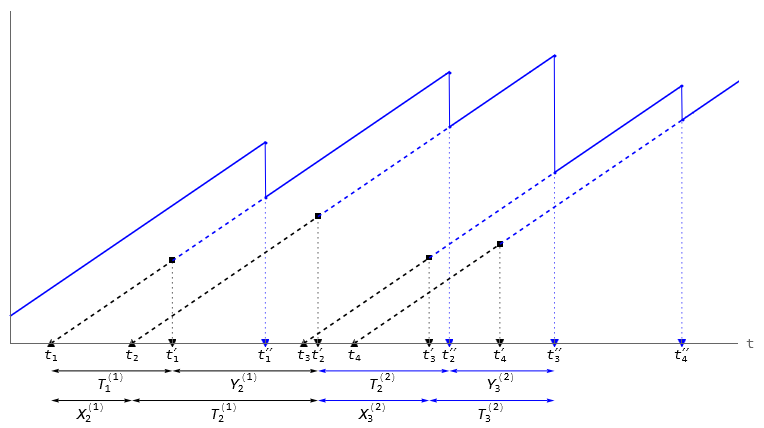}	
	\caption{Sample path for age in a tandem of two queues M/M/1/$\infty$ $\rightarrow$ M/M/1/$\infty$. Packets arriving at the second queue have a positive age equal to the system time in the first queue. Note that $Y_n^{(1)}$ are inter-departure times in the first queue and inter-arrival times in the second queue.}\label{fig::aoi_mm1_2}
\end{figure}

\autoref{fig::aoi_mm1_2} shows a sample path of the age at the end of this tandem. Arrivals at the first queue, represented by black triangles, occur at epochs $t_n$. After waiting in the queue if the server is busy upon arrival, the updates are served in a $\gamma$-exponential time, depart from the first queue and join the second at epochs $t'_n$, marked with inverted black triangles. If the second server is found busy, the update waits again until it starts service, where it spends a $\mu$-exponential time, leaving the system at epochs $t''_n$. The final departures are represented in \autoref{fig::aoi_mm1_2} by inverted blue triangles. Note that the designation of this tandem as M/M/1/$\infty$ $\rightarrow$ M/M/1/$\infty$ is completely justified, because it is well-known from Burke's Theorem \cite{burke:1956} that the departure process from a stable M/M/1$\infty$ queue --in this case, the first queue and thus, the arrival process at the second queue-- is also a Poisson process with rate $\lambda$. 

Using \autoref{thm::aaoi_au} to compute the AAoI at the end of this tandem yields:

\begin{equation}\label{eq::aaoi_tandem0}
	\Delta^A_{M/M/1/\infty} = \Delta^0_{M/M/1/\infty} + \dfrac{\E{Y_n^{(2)}A_{n-1}}}{\E{Y_n^{(2)}}},
\end{equation}
where $Y_n^{(2)}$ is the inter-departure time in the second queue, $A_n$ is the initial age and $\Delta^0_{M/M/1/\infty}$ is the average age in a M/M/1/$\infty$-FCFS queue with zero-age updates. It was found in \cite{kaul_real:2012} (see Equation (17)) that

\begin{equation}\label{eq::aaoi_mm1}
	\Delta^0_{M/M/1/\infty} = \inv{\lambda} + \inv{\mu} + \dfrac{\lambda^2/\mu^2}{\mu-\lambda}.
\end{equation}

Since there is no discarding in this tandem, all the packets go through both queues. Thus, upon arrival at the second queue, $A_n$ accounts for the time spent in the first queue, $A_n=T_n^{(1)}$. However, the successive application of the identity \eqref{eq::Yn_identity} on the inter-departure times from the second and first queues reveals that there is a statistical dependence between $Y_n^{(2)}$ and $T_{n-1}^{(1)}$, that needs to be addressed to apply \eqref{eq::aaoi_tandem0}:

\begin{align*}
	Y_n^{(2)} &= S_n^{(2)}+(X_n^{(2)}-T_{n-1}^{(2)})^+\\
	&= S_n^{(2)}+(Y_n^{(1)}-T_{n-1}^{(2)})^+\\
	&= S_n^{(2)}+\left(S_n^{(1)}+(X_n^{(1)}-T_{n-1}^{(1)})^+-T_{n-1}^{(2)}\right)^+.
\end{align*}

The following lemma, whose proof is provided in \autoref{app::appendix2}, solves this question.

\begin{lemma}\label{lemma::ET1Y2}
	\begin{equation*}
		\E{Y_n^{(2)}T_{n-1}^{(1)}} = \inv{\gamma\lambda}+\dfrac{\lambda/\gamma^2}{\gamma-\lambda}+\dfrac{\lambda/\gamma\mu}{\gamma+\mu-\lambda}
	\end{equation*}
\end{lemma}

Then, \autoref{thm::aaoi_tandem0} follows by applying \autoref{eq::aaoi_mm1}, \autoref{lemma::ET1Y2} and $\E{Y_n^{(2)}}=1/\lambda$ to \autoref{eq::aaoi_tandem0}.

\begin{remark}
	The very dynamics of the system prevent far updates from occurring in this type of tandems. Note that, since in this model $A_n=T_n^{(1)}$ and $X_n=X_n^{(2)}=Y_n^{(1)}$, the condition for an update to be obsolete is $T_n^{(1)}>Y_n^{(1)}+T_{n-1}^{(1)}$. However, by the identity \eqref{eq::Yn_identity} and the analogous one for the system time, $T_n=S_n+\left(T_{n-1}-X_n\right)^+$, it follows that the condition is equivalent to $$(T_{n-1}^{(1)}-X_n^{(1)})^+>(X_n^{(1)}-T_{n-1}^{(1)})^+-T_{n-1}^{(1)},$$ which is impossible if either $X_n^{(1)}>T_{n-1}^{(1)}$ or $X_n^{(1)}<T_{n-1}^{(1)}$.  
\end{remark}
\begin{remark}
	The procedure developed in this section can also derive the AAoI expressions for other well-known tandems. Specifically, the independence between final inter-departure times and initial ages can be exploited in tandems consisting of two single-capacity queues, with both preemption in service \cite[Theorem 2]{yates:2018} and without preemption in service \cite[Theorem 1]{kam:2022}.
\end{remark}
\subsubsection{$C+1$ M/M/1/$\infty$ queues in tandem}
As acknowledged in \cite{kam:2022}, while the rationale employed for the two-queued tandem of \autoref{sec::tandem2} can be extended for more than two queues, the practical computation of the AAoI in larger tandems involves a chain of conditional variables whose treatment is quite complex. This is where the result of \autoref{cor::interval} becomes particularly valuable, as it provides a straightforward interval for the AAoI of a tandem formed by any arbitrary number of M/M/1/$\infty$ queues.

Suppose that the service rate of the $i$-th server is $\gamma_i,\ i=1,\ldots,C+1$. For consistency with the previous section, the notation $\gamma_{C+1}=\mu$ is used interchangeably to denote the service rate of the last server. Conveniently, the notation $\Delta$ is used to denote the AAoI at the end of the tandem, and $\Delta_i$ to denote the AAoI of a single queue with service rate $\gamma_i$. By Burke's Theorem, the departure process at the end of such a tandem remains Poisson of rate $\lambda$, implying that the coefficient of variation of the inter-departure times is $\kappa_Y=1$. Additionally, the initial age of the updates entering the last queue is the sum of the system times in each of the preceding queues $$A_n=\sum\limits_{i=1}^CT_n^{(i)}.$$ Since each system time is exponentially distributed, $A_n$ is hypoexponential of parameters $(\gamma_1-\lambda,\ldots,\gamma_C-\lambda)$. Therefore, the mean and standard deviation of $A_n$ are given by 
\begin{equation*}
	\E{A_n}  = \sum\limits_{i=1}^C\inv{\gamma_i-\lambda}\quad \text{ and }\quad \sigma_A = \sqrt{\sum\limits_{i=1}^C\inv{(\gamma_i-\lambda)^2}},\\
\end{equation*} 
respectively. Then, by \autoref{cor::interval}, the AAoI at the end of the tandem is bounded within the interval

\begin{align}\label{eq::bound_aaoi_tandem}
	\Delta &\in \left[\Delta_{C+1}^0+\E{A_n} \pm \kappa_Y\sigma_A\right]\nonumber\\
	&=\left[\left(\inv{\lambda}+\inv{\mu}+\dfrac{\lambda^2/\mu^2}{\mu-\lambda}\right) + \sum\limits_{i=1}^C\inv{\gamma_i-\lambda} \pm \sqrt{\sum\limits_{i=1}^C\inv{(\gamma_i-\lambda)^2}} \right].
\end{align}  

Moreover, evidence suggests the AAoI at the end of this tandem is independent of the ordering of the servers \cite{kam:2022}. Although this hypothesis has not been formally proven, it is intuitively sensible as all arrival and departure processes are Poisson of rate $\lambda$, and the distribution of the time spent in a queue is the same regardless of its position in the tandem. Our numerical simulations strongly support this conjecture, as shown in the following tables. 
Tables \ref{tab::3_servers}, \ref{tab::6_servers} and \ref{tab::10_servers} present results for tandems of 3, 6, and 10 queues respectively, with a normalised arrival rate $\lambda=1$, equispaced loading rates ranging from $0.1$ to $0.9$ and different server orderings. For each configuration, we conducted 100 simulations using the \texttt{simmer} library \cite{ucar:2019} in \texttt{R} \cite{r_core:2022}. The columns \textit{age av} and \textit{age sd} show the mean and standard deviation of the AAoI across runs. The consistency of the empirical age across different orderings strongly supports the server-order independence conjecture.

\begin{table}[H]
	\begin{center}
		\caption{Simulated AAoI for a tandem of 3 queues, $\rho_i=1/\gamma_i$, all different orderings.}\label{tab::3_servers}
		\begin{tabular}{@{\hspace{1.2\tabcolsep}} c c c c c c c @{\hspace{1.2\tabcolsep}}}
			\toprule[1.5 pt]
			$\rho_1$&$\rho_2$&$\rho_3$& age av & age sd & age lb & age ub  \\
			\midrule[0.8 pt]
			\cg 0.1 &\cg0.5  &\cg0.9  &\cg10.1 &\cg1.53 &\cg9.29 &\cg11.3 \\
			0.1     &  0.9   &  0.5   &  9.86  &  1.40  &  1.86  & 19.9 \\
			\cg0.5  &\cg0.1  &\cg0.9  &\cg10.1 &\cg1.87 &\cg9.29 &\cg11.3 \\
			0.5     &  0.9   &  0.1   &  10.3  &  1.67  &  2.05  & 20.2 \\ 
			0.9     &  0.1   &  0.5   &  10.1  &  1.73  &  1.86  & 19.9 \\
			0.9     &  0.5   &  0.1   &  10.2  &  1.76  &  2.05  & 20.2 \\
			\bottomrule[1.5 pt]
		\end{tabular}
	\end{center}
\end{table}
\begin{table}[H]
	\begin{center}
		\caption{Simulated AAoI for a tandem of 6 queues, $\rho_i=1/\gamma_i$, 6 different orderings.}\label{tab::6_servers}
		\begin{tabular}{@{\hspace{1.2\tabcolsep}} c c c c c c c c c c @{\hspace{1.2\tabcolsep}}}
			\toprule[1.5 pt]
			$\rho_1$&$\rho_2$&$\rho_3$&$\rho_4$&$\rho_5$&$\rho_6$& age av & age sd & age lb & age ub  \\
			\midrule[0.8 pt]
			\cg0.10 &\cg0.26 &\cg0.42 &\cg0.58 &\cg0.74 &\cg0.90 &\cg14.4 &\cg1.79 &\cg11.3 &\cg17.9 \\
			0.90 & 0.74 & 0.58 & 0.42 & 0.26 & 0.10 & 14.4 & 1.95 & 5.83 & 25.0 \\
			0.26 & 0.10 & 0.90 & 0.58 & 0.74 & 0.42 & 14.3 & 2.01 & 5.69 & 24.8 \\
			0.10 & 0.74 & 0.26 & 0.58 & 0.90 & 0.42 & 14.5 & 1.72 & 5.69 & 24.8 \\ 
			0.74 & 0.58 & 0.90 & 0.10 & 0.42 & 0.26 & 14.5 & 1.50 & 5.78 & 24.9 \\
			0.10 & 0.90 & 0.42 & 0.26 & 0.74 & 0.58 & 14.4 & 1.52 & 5.60 & 24.6 \\
			\bottomrule[1.5 pt]
		\end{tabular}
	\end{center}
\end{table}
\newcolumntype{g}{>{\centering\arraybackslash}p{0.5cm}}
\newcolumntype{f}{>{\centering\arraybackslash}p{1.02cm}}
\begin{table}[H]
	\begin{center}
		\caption{Simulated AAoI for a tandem of 10 queues, 6 different orderings, $\rho_i=1/\gamma_i$.}\label{tab::10_servers}
		\begin{tabular}{@{\hspace{1.2\tabcolsep}} g g g g g g g g g g f f f f @{\hspace{1.2\tabcolsep}}}
			\toprule[1.5 pt]
			$\rho_1$&$\rho_2$&$\rho_3$&$\rho_4$&$\rho_5$&$\rho_6$&$\rho_7$&$\rho_8$&$\rho_9$&$\rho_{10}$& age av & age sd & age lb & age ub  \\
			\midrule[0.8 pt]
			\cg0.10 &\cg0.19&\cg0.28&\cg0.37&\cg0.46&\cg0.54&\cg0.63&\cg0.72&\cg0.81&\cg0.90 &\cg20.9 &\cg1.88 &\cg15.6 &\cg26.7\\
			0.90 & 0.81& 0.72& 0.63& 0.54& 0.46& 0.37& 0.28& 0.19& 0.10 & 21.0 & 2.06 & 11.4 & 32.5\\
			0.10 & 0.54& 0.37& 0.72& 0.81& 0.28& 0.90 & 0.19& 0.46& 0.63& 21.1 & 2.12 & 11.1 & 32.0\\
			0.37& 0.46& 0.10 & 0.90 & 0.63& 0.28& 0.19& 0.72& 0.54& 0.81& 20.5 & 1.84 & 11.6 & 31.0\\ 
			0.81& 0.54& 0.37& 0.46& 0.72& 0.28& 0.90 & 0.19& 0.63& 0.10 & 20.8 & 1.84 & 11.4 & 32.5\\
			0.90 & 0.28& 0.37& 0.10 & 0.54& 0.19& 0.63& 0.72& 0.46& 0.81& 20.8 & 1.91 & 11.6 & 31.0\\
			\bottomrule[1.5 pt]
		\end{tabular}
	\end{center}
\end{table}

Furthermore, these tables reveal a clear pattern for the bounds from \autoref{eq::bound_aaoi_tandem}: the tightest bounds are consistently achieved when the slowest server is placed last (shaded rows). This occurs because the standard deviation of the initial age $\sigma_A$, and therefore the interval width, is minimised when the slowest server is at the end of the tandem. This leads to a significant practical implication: if the server-order independence holds, then the tightest possible bounds for any tandem configuration are given by the bounds calculated for the specific order where the slowest server is last. Therefore, regardless of the actual system configuration, a practitioner can always obtain the most accurate interval estimate by simply assuming the slowest server is in the final position when applying \autoref{cor::interval}.

In all simulations, Tables \ref{tab::3_servers}, \ref{tab::6_servers} and \ref{tab::10_servers} show that the empirical AAoI falls within the theoretical bounds --collected in the columns \textit{age lb} and \textit{age ub}--, confirming their validity. However, the bounds can be loose (50-100\% error) for unfavourable server orderings, which underscores the importance of the aforementioned ordering strategy to obtain the best possible estimates from our framework.

Furthermore, a closer look at the table reveals that the empirical age lies near the centre of the interval, always closer to the lower bound. This is not coincidental, but a direct consequence of the dependence structure in this tandem; specifically, a systematic negative correlation ($r_{Y,A}<0$) between the initial age $A_{n-1}$ (which is the total system time in previous queues) and the final inter-departure time $Y_n$.

The correlation is negative due to the dynamics captured by the Lindley-type equation for the inter-departure times. Expanding the equation of the inter-departure time in the last queue, $Y_n^{(C+1)}$, reveals
\begin{align*}
	Y_n^{(C+1)} &= T_n^{(C+1)} + X_n^{(C+1)} - T_{n-1}^{(C+1)}\\
	&= T_n^{(C+1)} + Y_n^{(C)} - T_{n-1}^{(C+1)}\\
	&= T_n^{(C+1)} + S_n^{(C)} + (X_n^{(C)}-T_{n-1}^{(C)})^+- T_{n-1}^{(C+1)},
\end{align*}
where $A_{n-1}$ is a function of $T_{n-1}^{(C)}$ and previous system times. When $T_{n-1}^{(C)}$, a component of $A_{n-1}$, is large, the term $(X_n^{(C)}-T_{n-1}^{(C)})^+$ tends to zero. This suppresses the idle time, making $Y_n^{(C+1)}$ smaller. Conversely, when $T_{n-1}^{(C)}$ is small, the idle term is more likely to be positive, increasing $Y_n^{(C+1)}$. This inverse relationship, propagated through the tandem, results in the observed negative correlation $r_{Y,A}<0$. For example, using \autoref{lemma::ET1Y2}, the covariance between $Y_n^{(C+1)}$ and $T_{n-1}^{(C)}$ can be explicitly computed as $$\Cov{Y_n^{(C+1)},T_{n-1}^{(C)}}=\dfrac{1}{\gamma^2_C}\dfrac{(\gamma_C+\gamma_{C+1})(\lambda-\gamma_{C+1})}{\gamma_{C+1}(\gamma_C+\gamma_{C+1}-\lambda)},$$ which is indeed negative for stable queues ($\gamma_i>\lambda$). Since the correction term is $\E{A_n} + r_{Y,A}\cdot\kappa_Y\cdot\sigma_A$, a negative correlation subtracts from the mean age, pulling the final AAoI down from the upper bound towards the lower bound.

While a precise calculation of $r_{Y,A}$ is complex, its consistent negativity suggests a practical rule of thumb for estimating the correction term in this specific tandem. As the AAoI often lies in the lower half of the bounding interval, a simple yet effective approximation is to use a weighted average leaning towards the lower bound (e.g., $\E{Y_nA_{n-1}}\approx0.55\cdot LB + 0.45\cdot UB$) for a first-order estimate. Likewise, in practice, the upper limit of the interval can be set at $\E{A_n}$.

In the specific case of homogeneous servers, $\gamma_i=\mu,\ \forall i=1,\ldots,C+1$, \autoref{eq::bound_aaoi_tandem} reduces to 
\begin{equation*}
	\Delta\in\left[\left(\inv{\lambda} + \inv{\mu} + \dfrac{\lambda^2/\mu^2}{\mu-\lambda}\right) + \dfrac{C}{\mu-\lambda} \pm \dfrac{\sqrt{C}}{\mu-\lambda}\right].
\end{equation*}
that gives and interval of width $2\sqrt{C}/(\mu-\lambda)$. It can be observed that, for faster servers (i.e., larger $\mu$), both limits move to the left and the width becomes tighter, while for a larger number of servers, the interval moves to the right and gets wider.
\subsubsection{M/M/1/1 $\rightarrow$ $\cdot$/M/1/$\infty$ tandem. A case study with heterogeneous queues}
We now turn to a system that, to the best of our knowledge, has not yet been analytically characterised from an age perspective: a tandem of two heterogeneous queues. This example will highlight the framework's ability to derive new analytical results.

Consider a tandem formed of two different queues. Updates arrive at the first queue in a Poisson Process of rate $\lambda$. If the server is idle, the update is processed in a $\gamma$-rate exponential time, but packets that arrive while the server is busy are dropped. Once the service is completed in this first queue, the packet enters a second infinite-capacity queue, in which packets are processed in order of arrival in a $\mu$-rate exponential time. This can be referred to as M/M/1/1$\rightarrow$HE/M/1/$\infty$ tandem because the departures from the first queue --and, thus, the arrivals at the second-- follow a hypoexponential distribution of parameters $(\lambda,\gamma$) \cite[Section IV.A]{costa:2016}.

In this model, \autoref{thm::aaoi_au} applied to the calculation of the AAoI at the end of the tandem results in

\begin{equation}\label{eq::aaoi_tandem2}
	\Delta_{M/M/1/1\rightarrow HE/M/1/\infty} = \Delta^0_{HE/M/1/\infty} + \dfrac{\E{Y_nA_{n-1}}}{\E{Y_n}},
\end{equation}
where $Y_n$ is the inter-departure time of the second queue, $A_n$ is the age of the $n$th update when entering the second queue, and $\Delta^0_{HE/M/1/\infty}$ is the AAoI with zero-age updates defined in the following Proposition, whose demonstration is presented in Appendix \ref{app::appendix4}. 

\begin{proposition}\label{prop::aaoi_hem1}
	The AAoI of the HE/M/1/$\infty$ queueing model is given by 
	
	\begin{equation*}
		\Delta_{HE/M/1/\infty}^0 = \dfrac{1}{\lambda} + \dfrac{1}{\gamma} + \dfrac{1}{\mu} +\dfrac{\sigma\rho}{\mu-\mu\sigma} - \dfrac{1-\sigma^2}{\lambda+\gamma}.
	\end{equation*}
	where $\rho=\E{S}/\E{X}=\dfrac{\lambda\gamma}{\mu(\lambda+\gamma)}<1$ and $\sigma$ is the unique root in $(0,1)$ of the equation $\sigma=\tilde X(\mu-\mu\sigma)$, with $\tilde X(\cdot)$ the Laplace-Stieltjes Transform of the inter-arrival time distribution.
\end{proposition}

In this tandem, all packets departing from the first queue are received and served in the second queue. Therefore, $A_n$ is the system time of the first queue, which is an exponentially distributed service time of rate $\gamma$ \cite[Section IV.A]{costa:2016}, and independent of the inter-departure time of future packets in the second queue. Thus, the second term in \autoref{eq::aaoi_tandem2} reduces to $\E{A_{n-1}}=1/\gamma$, and the age at the end of this tandem results in

\begin{align*}
	\Delta_{M/M/1/1\rightarrow HE/M/1/\infty} &= \left(\dfrac{1}{\lambda} + \dfrac{1}{\gamma} + \dfrac{1}{\mu} +\dfrac{\sigma\rho}{\mu-\mu\sigma} - \dfrac{1-\sigma^2}{\lambda+\gamma}\right) + \E{A_{n-1}} \nonumber\\
	&= \dfrac{1}{\lambda} + \dfrac{2}{\gamma} + \dfrac{1}{\mu} +\dfrac{\sigma\rho}{\mu-\mu\sigma} - \dfrac{1-\sigma^2}{\lambda+\gamma}
\end{align*}

\begin{remark}
	This tandem also inherently prevents the arrival of far updates. Note that, in this case, $A_n=S_n^{(1)}$ and $X_n=X_n^{(2)}=Y_n^{(1)}$. Using again identity \eqref{eq::Yn_identity}, the condition for an obsolete packet becomes $0>(X_n^{(1)}-S_{n-1}^{(1)})^++S_{n-1}^{(1)}$, which is also impossible.
\end{remark}
\subsection{Retrial queues: A novel application}
Retrial queues represent a classic yet complex model in queueing theory. However, their analysis in the context of AoI remains an open challenge. In this section, we leverage the aged-updates framework to provide the first characterisation of the AAoI in an M/M/1 retrial queue.

\begin{wrapfigure}{r}{0.5\textwidth}
	\centering
	\includegraphics[width=0.45\textwidth]{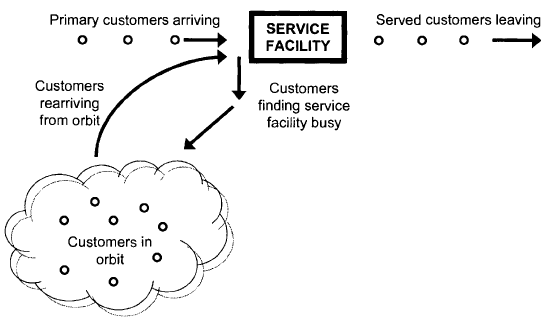}	
	\caption{Basic model for a retrial queue (from \cite[Section 3.5]{gross:fqt}).}\label{fig::retrial_queue}
\end{wrapfigure} 

A \emph{retrial queue} is a type of queueing model where an arrival from the source —referred to as a \emph{primary arrival}— temporarily leaves the system when it finds the server busy and returns later. While waiting to re-enter, the packet is said to be in an \emph{orbit}. When the server becomes available, both the source and the orbit compete for access, each submitting packets based on its own arrival process. The system is sketched in \autoref{fig::retrial_queue}. There are two key differences between a standard queue and the orbit in retrial queues. First, packets in the orbit cannot monitor the server's state; they can only check their status by attempting to re-enter the system, an event known as a \emph{retrial}. Furthermore, after a service is completed, there is a period during which the server is idle until a new update arrives, whether it be a primary arrival or a packet from the orbit. Secondly, to better reflect the real-life systems it is intended to represent (such as customers repeatedly calling a store until they receive service), packets in the orbit generally do not follow any specific order; thus, the service discipline is random.

Retrial queueing models are generally difficult to analyse, and the model described here is one of the few that can be solved analytically. Therefore, it is not surprising that, to the best of our knowledge, the AoI has not yet been studied in this class of models. However, using the aged-updates scheme, the calculation of AAoI can be significantly simplified, enabling the derivation of either an exact expression or reasonable bounds. For clarity, the specific model considered in this example will first be described, followed by an explanation of how to apply the scheme of aged updates to determine the AAoI.
\subsubsection{Retrial queueing model example}
Consider a M/M/1 retrial queue with single re-attempts\footnote{In contrast to the model studied in \cite[Section 3.5.1]{gross:fqt}, where multiple retrials occur.}. Primary arrivals and retrials follow Poisson processes with rates $\lambda$ and $\theta$, respectively. Any primary arrival that finds a busy server joins the orbit. Similarly, if an update attempts to re-enter the system from the orbit and finds the server busy, it returns to the orbit. When the server is idle, the first arrival —whether it is a primary arrival or a retry from the orbit— enters service immediately. The processed updates are delivered at exponential times of rate $\mu$. 

\begin{figure}[htpb]
	\centering
	\includegraphics[width=0.7\textwidth, clip=true, trim=2.39cm 12.97cm 10.1cm 1.95cm]{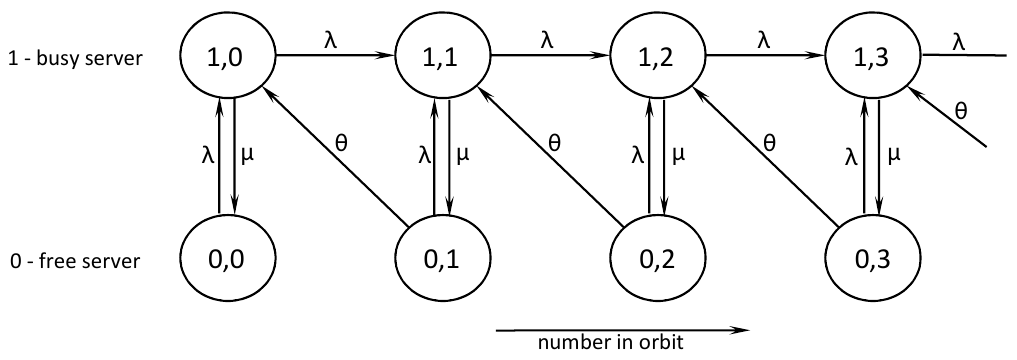}	
	\caption{\footnotesize{Flow diagram for the retrial queue with single retrials. Arrivals at the server occur at rates $\lambda$ --fresh, from the source-- or $\theta$ --aged, from the orbit--, and departures from the server occur at exponential times of rate $\mu$.}}\label{fig::retrial_diag}
\end{figure}

This system can be modelled using a Continuous Time Markov Chain with state space $\left\{(i,n),i\in\{0,1\},\ n\in\mathbb Z^+\right\}$, where $i$ represents the server state and $n$ is the number of updates in orbit. \autoref{fig::retrial_diag} shows the transitions and rates for this Markov Chain.   

Let $p_{i,n}$ denote the steady-state probability that the system is in state $(i,n)$. The following theorem provides the steady-state solutions for the described model.

\begin{theorem}\label{thm::rq_solution}
	Let $\rho=\lambda/\mu$ denote the server utilisation rate and $\pi=\theta/(\lambda+\theta)$ the probability that the in-service packet comes from orbit. The steady-state probabilities for the system above are given by
	\begin{subequations}\label{eq::retrial_sp}
		\begin{align}
			p_{0,0} &= 1-\rho/\pi\label{eq::p00}\\
			p_{0,n} &= (1-\rho/\pi)(1-\pi)(\rho/\pi)^n,\qquad n\geq 1\label{eq::p0n}\\
			p_{1,n} &= (1-\rho/\pi)\rho(\rho/\pi)^n,\qquad n\geq 0\label{eq::p1n}
		\end{align}
	\end{subequations}
\end{theorem}

\begin{myProof}
	See \autoref{app::appendix4}.
\end{myProof}

\begin{remark}
	Notably, the probability of finding the server busy in this system proves to be independent of the retrial rate $\theta$, and the same as in a standard M/M/1/$\infty$ model:
	
	\begin{equation*}
		\Pb{\text{busy server}} = \sum\limits_{n\geq0} p_{1,n} = (1-\rho/\pi)\rho\sum\limits_{n\geq0}(\rho/\pi)^n = \rho
	\end{equation*}
	Moreover, in the limiting case of instantaneous retrials (i.e., $\theta\to\infty$ and, therefore, $\pi\to1$) the steady-state probabilities in \eqref{eq::retrial_sp} approach those of the M/M/1/$\infty$ queue. In that case, the orbit behaves like an ordinary queue, and the random service discipline in the orbit does not affect the long-term state of the system. It is also worth mentioning that, for the probabilities of \autoref{thm::rq_solution} to make sense, it is necessary that $\rho<\pi$, which is stronger than the usual stability condition $\rho<1$. Intuitively, this inequality is equivalent to requiring that the retrial rate is high enough to ensure that the orbit does not grow infinitely.
\end{remark}

Let $L_o$ be the mean number of updates in orbit and $W_o$ the mean time spent in orbit. $L_o$ can be obtained using the steady-state probabilities \eqref{eq::retrial_sp}:
\begin{align}
	L_o &= \sum\limits_{n=1}^\infty n(p_{0,n} + p_{1,n})\nonumber\\
	&= (1-\rho/\pi)(1+\rho-\pi)\sum\limits_{n=1}^\infty n(\rho/\pi)^n\nonumber\\
	&= (1-\rho/\pi)(1+\rho-\pi)\sum\limits_{i=1}^\infty\left(\sum\limits_{n=1}^\infty (\rho/\pi)^n\right)\nonumber\\
	&= \dfrac{\rho(1+\rho-\pi)}{\pi-\rho}\label{eq::Lo}.
\end{align}

In turn, $W_o$ follows from Little's Law \cite{little:1961} applied to the orbit,
\begin{equation}\label{eq::Wo}
	W_o = L_o/\lambda = \dfrac{1+\rho-\pi}{\mu(\pi-\rho)}.
\end{equation}
\subsubsection{AAoI using the aged-updates scheme}
\begin{figure}[htpb]\captionsetup[subfigure]{font=footnotesize}
	\centering
	\subfloat[]{\includegraphics[width=0.48\textwidth]{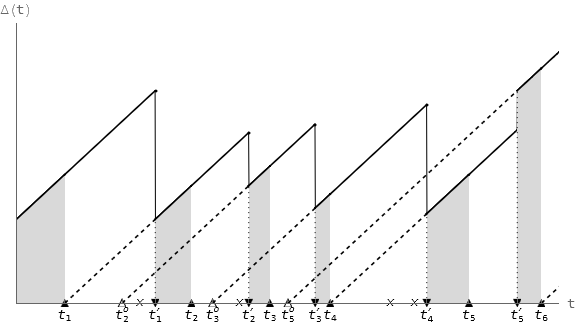}}\qquad
	\subfloat[]{\includegraphics[width=0.48\textwidth]{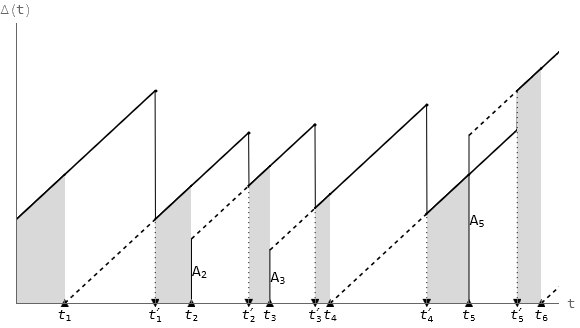}}
	\caption{Retrial queue with single re-attempts as a system with aged updates. (a) Retrial queueing standard model. (b) Equivalent M/M/1/1 queue without pre-emption and with initial age process $A_n$. Primary arrivals in the retrial queue are fresh updates, while arrivals from orbit are updates with a positive initial age.}\label{fig::rq_model}
\end{figure}

\autoref{fig::rq_model}a) shows an example of a sample path for the age in the M/M/1 retrial queue. Arrivals at the server, represented by black triangles, occur at epochs $t_n$, while departures are marked with inverted black triangles at epochs $t'_n$. Primary arrivals go directly to the server, while updates arriving at epochs $t^o_n$ find the server busy and join the orbit, starting service after possibly several re-attempts --denoted by crosses in \autoref{fig::rq_model}(a)--. After departures, the shaded intervals represent the period during which the server remains idle until the next arrival.

\autoref{fig::rq_model}b) depicts the age evolution in an analogous system in which the updates coming from orbit can be interpreted as carrying a positive initial age $A_n$ equivalent to the waiting time in orbit. The analogous system can be modelled as a single-capacity M/M/1/1 queue without pre-emption in service and arrival rate $\lambda+\theta$.

The AAoI in the M/M/1 retrial queueing model can be obtained applying \autoref{thm::aaoi_au} to the equivalent M/M/1/1 queue without pre-emption in service of \autoref{fig::rq_model}b),

\begin{equation}\label{eq::aaoi_rq}
	\Delta_{RQ} = \Delta_{RQ}^0 + \dfrac{\E{Y_n A_{n-1}}}{\E{Y_n}}.
\end{equation} 

The first term in the right-hand side of \eqref{eq::aaoi_rq} is the AAoI in a M/M/1/1 queue without pre-emption in service. By \cite[Equation (21)]{costa:2016}, this expression is 

\begin{equation}\label{eq::aaoi_mm11}
	\Delta_{RQ}^0 = \inv{\lambda+\theta} + \inv{\mu} + \dfrac{\lambda+\theta}{\mu(\lambda+\theta+\mu)}.
\end{equation}

The age process $A_n$ corresponds to the waiting time in orbit of the incoming updates, which is zero for arrivals from the source and positive for arrivals from the orbit. The random service discipline in orbit means that the waiting time of a packet does not influence when it is selected for service. Once a successful retry occurs, the service is exponential and therefore memoryless. These two facts ensure that waiting times are independent of future inter-departure times, resulting in the simplified case presented in \autoref{sec::independency}. Using \eqref{eq::Wo}, the average initial age is then given by 

\begin{align}
	\E{A_n} &= \Pb{\text{update from orbit}}\times\text{mean time in orbit} = \pi W^o\nonumber\\
	&= \dfrac{\pi(1+\rho-\pi)}{\mu(\pi-\rho)}\label{eq::EA_rq}.
\end{align} 

Plugging \eqref{eq::aaoi_mm11} and \eqref{eq::EA_rq} into \eqref{eq::aaoi_rq}, the AAoI in the M/M/1 retrial queue results in 

\begin{align*}
	\Delta_{RQ} &= \inv{\lambda+\theta} + \inv{\mu} + \dfrac{\lambda+\theta}{\mu(\lambda+\theta+\mu)} + \dfrac{\pi(1+\rho-\pi)}{\mu(\pi-\rho)} \\
	&= \inv{\mu}\left[\dfrac{1+\rho-\pi}{\rho} + \dfrac{\rho}{1+\rho-\pi} + \dfrac{\pi(1+\rho-\pi)}{\pi-\rho}\right].
\end{align*}
\section{Conclusions}\label{sec::conclusions}
This work introduced a general framework for analysing the Age of Information (AoI) in systems where updates arrive with a non-zero initial age, a frequent scenario in multi-hop networks, edge computing, and retransmission protocols. By relaxing the standard assumption of zero-age updates, we developed a more realistic model that captures the impact of prior delays.

The main contribution is a closed-form expression for the AAoI in single-server queues with aged updates. We demonstrate that the average age in this system equals the standard AAoI plus a correction term that captures the interaction between the packet's initial age and the inter-departure times. This formulation serves as a versatile analytical tool that bridges a significant gap in the literature regarding the computation of AAoI in more complex models. We examine its usefulness and potential advantages through comprehensive applications: recovering known results for systems like forwarding queues and homogeneous tandems under a unified approach; deriving the first AAoI characterization for novel models including the heterogeneous M/M/1/1 $\rightarrow$ $\cdot$/M/1/$\infty$ tandem and the M/M/1 retrial queue; and highlighting its practical relevance by establishing analytical bounds for systems with complex dependency structures, validated through consistent simulations of multi-queue tandems. The framework's flexibility regarding packet management policies, as formally established in this work, further enhances its applicability across diverse system configurations.

Several promising research directions emerge from this work. One natural extension involves applying the framework to multi-source systems where aged updates compete for server resources. Another direction would explore systems with non-i.i.d. arrival processes, where the correlation structure between queueing times and initial ages introduces additional complexity. The framework also provides a foundation for designing and optimising  optimal control policies that exploit initial age information, such as scheduling disciplines that prioritise fresher packets or admission control policies that strategically discard stale updates. The analysis could be extended to consider negative initial ages for systems where updates remain valid for a period before ageing, though this would likely require revisiting the graphical method used in our proof. Another promising direction involves applying the aged-updates framework to alternative age metrics like the Age of Incorrect Information or Query AoI. Finally, empirical validation through real-world measurements would help assess the practical accuracy of the derived expressions and bounds.
\section*{Code Availability Statement}
The code to reproduce the results in Tables \ref{tab::3_servers}, \ref{tab::6_servers} and \ref{tab::10_servers} is available at the repository

\begin{center}
	\href{https://github.com/spatialstatisticsupna/MTF-industrial-process}{https://github.com/spatialstatisticsupna/On-the-AoI-in-SSQ-with-Aged-Updates-Tables}.
\end{center}
\section*{Acknowledgements}
This research was supported in part by the Public University of Navarre under the grant for International Mobility for Doctoral Students, by the Department of Education of the Basque Government through the Consolidated Research Group MATHMODE (IT1456-22), by the ANR LabEx CIMI (grant ANR-11-LABX-0040) within the French State Programme ''Investissements d’Avenir'', and by the ANR under the France 2030 program, grant NF-NAI: ANR-22-PEFT-0003.
\bibliographystyle{IEEEtran}
\bibliography{refs}
\appendix
\section{Proof of Lemma \ref{lemma::ET1Y2}}\label{app::appendix2}
\begin{lemma}\label{lemma::3}
	Let $X\sim\expo{\lambda}$ and $Y$ be a non-negative rv with a general distribution. Then
	\begin{equation*}
		\Pb{X>Y}=\tilde Y(\lambda),
	\end{equation*}
	where $\tilde Y(s)=\E{\ex^{-sY}}$ is the Laplace-Stieltjes Transform of $Y$.
\end{lemma}
\begin{myProof}
	\begin{align*}
		\Pb{X>Y} &= \int_0^\infty\Pb{X>y}f_Y(y)\dd y = \int_0^\infty\ex^{-\lambda y}f_Y(y)\dd y = \E{\ex^{-\lambda Y}}\qedsymbol
	\end{align*}
\end{myProof}

The following lemma follows from elementary properties of exponential random variables.

\begin{lemma}\label{lemma::5}
	Let $X_1,X_2$ be independent exponential random variables of rate $\lambda_i$, and $Z=X_2-X_1$.

	\begin{enumerate}
		\item [a)] $\Pb{X_2>X_1}=\lambda_1/(\lambda_1+\lambda_2)$ (exponential version of Lemma \ref{lemma::3})
		\item [b)] Define the event $\Lambda=\{X_2>X_1\}$. Given $\Lambda$, $X_1$ and $Z$ are independent, and
		\begin{align*}
			X_1|\Lambda &\sim\expo{\lambda_1+\lambda_2}\\
			Z|\Lambda   &\sim\expo{\lambda_2}
		\end{align*}
		\item [c)] Given $\Lambda$, $X_1$ and $X_2$ are not independent, and $$X_2|\Lambda\sim\Hexp{\lambda_1,\lambda_1+\lambda_2}.$$
	\end{enumerate}
\end{lemma}

The above results are somewhat extended in the following lemma.

\begin{lemma}\label{lemma::4}
	Let $X\sim\expo{\lambda}$ and $Y\sim\Hexp{\mu_1,\mu_2}$ independent random variables, $Z=Y-X$, and define the event $\Lambda=\left\{Y>X\right\}$. If $$w=\dfrac{\mu_2(\lambda+\mu_2)}{(\mu_2-\mu_1)(\lambda+\mu_1+\mu_2)},$$
	then

	\begin{enumerate}
		\item[a)] $\Pb{Y>X}=\dfrac{\lambda(\lambda+\mu_1+\mu_2)}{(\lambda+\mu_1)(\lambda+\mu_2)}$
		\item[b)] the pdf of $Z|\Lambda$ is given by 
		\begin{equation*}
			f_{Z|\Lambda}(z) = w\mu_1\ex^{-\mu_1 z} + (1-w)\mu_2\ex^{-\mu_2 z},
		\end{equation*}

		\item[c)] the pdf of $X|\Lambda$ is

		\begin{equation*}
			f_{X|\Lambda}(x)=w(\lambda+\mu_1)\ex^{-(\lambda+\mu_1)x}+(1-w)(\lambda+\mu_2)\ex^{-(\lambda+\mu_2)x},
		\end{equation*}

		\item[d)] Given $\Lambda$, $Z$ and $X$ are not independent, and their joint pdf is given by

		\begin{equation*}
			f_{Z,X|\Lambda}(z,x) = w\mu_1\ex^{-\mu_1 z}(\lambda+\mu_1)\ex^{-(\lambda+\mu_1)x} + (1-w)\mu_2\ex^{-\mu_2 z}(\lambda+\mu_2)\ex^{-(\lambda+\mu_2)x}.
		\end{equation*}
	\end{enumerate}

\end{lemma}
\begin{myProof}

	\begin{enumerate}
		\item[a)] By Lemma \ref{lemma::3} as applied to hypoexponential distribution.
		\item[b)]

		\begin{align*}
			\Pb{Y-X>z|\Lambda} &= \left(\Pb{Y>X}\right)^{-1}\Pb{Y>X+z}\\
			&= \dfrac{(\lambda+\mu_1)(\lambda+\mu_2)}{\lambda(\lambda+\mu_1+\mu_2)}\int_0^\infty\Pb{Y>x+z}\Pb{X=x}\dd x\\
			&= \dfrac{(\lambda+\mu_1)(\lambda+\mu_2)}{\lambda+\mu_1+\mu_2}\left(\dfrac{\mu_2}{\mu_2-\mu_1}\ex^{-\mu_1 z}\int_0^\infty\ex^{-(\lambda+\mu_1)x}\dd x - \dfrac{\mu_1}{\mu_2-\mu_1}\ex^{-\mu_2 z}\int_0^\infty\ex^{-(\lambda+\mu_2)x}\dd x \right)\\
			&=\dfrac{\mu_2(\lambda+\mu_2)}{(\mu_2-\mu_1)(\lambda+\mu_1+\mu_2)}\ex^{-\mu_1 z}-\dfrac{\mu_1(\lambda+\mu_1)}{(\mu_2-\mu_1)(\lambda+\mu_1+\mu_2)}\ex^{-\mu_2 z},\\
			\intertext{which, after differentiation, yields Lemma \ref{lemma::4}b).}
		\end{align*}

		\item[c)]

		\begin{align*}
			\Pb{X>x|\Lambda} &= \left(\Pb{Y>X}\right)^{-1}\Pb{X>x,Y>X}\\
			&= \dfrac{(\lambda+\mu_1)(\lambda+\mu_2)}{\lambda(\lambda+\mu_1+\mu_2)}\int_x^\infty\Pb{Y>x}\Pb{X=x}\dd x\\	
			&= \dfrac{(\lambda+\mu_1)(\lambda+\mu_2)}{\lambda+\mu_1+\mu_2}\left(\dfrac{\mu_2}{\mu_2-\mu_1}\int_x^\infty\ex^{-(\lambda+\mu_1)x}\dd x - \dfrac{\mu_1}{\mu_2-\mu_1}\int_x^\infty\ex^{-(\lambda+\mu_2)x}\dd x \right)\\
			&=\dfrac{\mu_2(\lambda+\mu_2)}{(\mu_2-\mu_1)(\lambda+\mu_1+\mu_2)}\ex^{-(\lambda+\mu_1) x}-\dfrac{\mu_1(\lambda+\mu_1)}{(\mu_2-\mu_1)(\lambda+\mu_1+\mu_2)}\ex^{-(\lambda+\mu_2) x},\\
			\intertext{which, after differentiation, yields Lemma \ref{lemma::4}c).}
		\end{align*}

		\item[d)]

		\begin{align*}
			\Pb{Y-X>z,X>x|\Lambda} &= \left(\Pb{Y>X}\right)^{-1}\Pb{Y>X+z,X>x}\\
			&= \dfrac{(\lambda+\mu_1)(\lambda+\mu_2)}{\lambda(\lambda+\mu_1+\mu_2)}\int_x^\infty\Pb{Y>t+z}\Pb{X=t}\dd t\\
			&= \dfrac{(\lambda+\mu_1)(\lambda+\mu_2)}{\lambda+\mu_1+\mu_2}\left(\dfrac{\mu_2}{\mu_2-\mu_1}\ex^{-\mu_1 z}\int_x^\infty\ex^{-(\lambda+\mu_1)t}\dd t - \dfrac{\mu_1}{\mu_2-\mu_1}\ex^{-\mu_2 z}\int_x^\infty\ex^{-(\lambda+\mu_2)t}\dd t \right),\\
			\intertext{which, after differentiation, yields Lemma \ref{lemma::4}d). The dependency follows from the inequality}
			& f_Z(z)f_{X|\Lambda}(x)\ne f_{Y-X,X|\Lambda}(z,x)
		\end{align*}

		$\qedsymbol$
	\end{enumerate}

\end{myProof}
\begin{myProof}[Lemma \ref{lemma::ET1Y2}]
	Using identity \eqref{eq::Yn_identity} with $Y_n^{(2)}$ and $Y_n^{(1)}$, the inter-departure times in the second queue can be decomposed as $$Y_n^{(2)}=S_n^{(2)}+\left(S_n^{(1)}+(X_n^{(1)}-T_{n-1}^{(1)})^+-T_{n-1}^{(2)}\right)^+.$$ Recall that the system times in both queues are exponential, of rate $(\gamma-\lambda)$ in the first queue and rate $(\mu-\lambda)$ in the second one. Conditioning on the possible outcomes of the differences yields

	\begin{align*}
		Y_n^{(2)}T_{n-1}^{(1)} &= S_n^{(2)}T_{n-1}^{(1)} + 
		\begin{cases}
			\left(S_n^{(1)}-T_{n-1}^{(2)}\right)^+T_{n-1}^{(1)}&\text{if }X_n^{(1)}<T_{n-1}^{(1)}\\
			\left(S_n^{(1)}+(X_n^{(1)}-T_{n-1}^{(1)})-T_{n-1}^{(2)}\right)^+T_{n-1}^{(1)}&\text{if }X_n^{(1)}>T_{n-1}^{(1)}
		\end{cases}\\
		&=  S_n^{(2)}T_{n-1}^{(1)} + 
		\begin{cases}
			\left(S_n^{(1)}-T_{n-1}^{(2)}\right)T_{n-1}^{(1)}&\text{if }X_n^{(1)}<T_{n-1}^{(1)}\text{ and }S_n^{(1)}>T_{n-1}^{(2)}\\
			\left(S_n^{(1)}+(X_n^{(1)}-T_{n-1}^{(1)})-T_{n-1}^{(2)}\right)T_{n-1}^{(1)}&\text{if }X_n^{(1)}>T_{n-1}^{(1)}\text{ and }S_n^{(1)}+(X_n^{(1)}-T_{n-1}^{(1)})>T_{n-1}^{(2)}
		\end{cases}
	\end{align*} 

	For the first case, by Lemma \ref{lemma::5}b),c)

	\begin{align*}
		(S_n^{(1)}-T_{n-1}^{(2)})|S_n^{(1)}>T_{n-1}^{(2)}&\sim\expo{\gamma}\\
		T_{n-1}^{(1)}|T_{n-1}^{(1)}>X_n^{(1)} &\sim\Hexp{\gamma-\lambda,\gamma},
	\end{align*}
	being also independent of each other. For the second case, 

	\begin{align*}
		(X_n^{(1)}-T_{n-1}^{(1)})|X_n^{(1)}>T_{n-1}^{(1)} &\sim\expo{\lambda}\text{, by Lemma \ref{lemma::5}b)}\\
		S_n^{(1)}+(X_n^{(1)}-T_{n-1}^{(1)})|X_n^{(1)}>T_{n-1}^{(1)} &\sim\Hexp{\gamma,\lambda}\text{, by definition}\\
		S_n^{(1)}+(X_n^{(1)}-T_{n-1}^{(1)})-T_{n-1}^{(2)}|S_n^{(1)}+(X_n^{(1)}-T_{n-1}^{(1)})>T_{n-1}^{(2)},X_n^{(1)}>T_{n-1}^{(1)}&\sim w\cdot\expo{\lambda}+(1-w)\cdot\expo{\gamma}\text{, by Lemma \ref{lemma::4}a)}\\
		T_{n-1}^{(1)}|X_n^{(1)}>T_{n-1}^{(1)}&\sim\expo{\gamma}\text{, by Lemma \ref{lemma::5}b)},
	\end{align*}
	where $$w=\dfrac{\gamma(\gamma+\mu-\lambda)}{(\gamma-\lambda)(\gamma+\mu)}.$$ The independence between the last two also follows from Lemma \ref{lemma::5}b). All this results in

	\begin{align*}
		\E{Y_n^{(2)}T_{n-1}^{(1)}} =& \ \E{S_n^{(2)}}\E{T_{n-1}^{(1)}}\\
		&+ \E{(S_n^{(1)}-T_{n-1}^{(2)})|S_n^{(1)}>T_{n-1}^{(2)}}\E{T_{n-1}^{(1)}|T_{n-1}^{(1)}>X_n^{(1)}}\Pb{S_n^{(1)}>T_{n-1}^{(2)}}\Pb{T_{n-1}^{(1)}>X_n^{(1)}}\\
		&+ \E{S_n^{(1)}+(X_n^{(1)}-T_{n-1}^{(1)})-T_{n-1}^{(2)}|S_n^{(1)}+(X_n^{(1)}-T_{n-1}^{(1)})>T_{n-1}^{(2)},X_n^{(1)}>T_{n-1}^{(1)}}\\
		&\quad \cdot\E{T_{n-1}^{(1)}|X_n^{(1)}>T_{n-1}^{(1)}}\Pb{S_n^{(1)}+(X_n^{(1)}-T_{n-1}^{(1)})>T_{n-1}^{(2)}|X_n^{(1)}>T_{n-1}^{(1)}}\Pb{X_n^{(1)}>T_{n-1}^{(1)}}\\
		=&\inv{\mu}\inv{\gamma-\lambda}+\inv{\gamma}\left(\inv{\gamma-\lambda}+\inv{\gamma}\right)\dfrac{\mu-\lambda}{\gamma+\mu-\lambda}\dfrac{\lambda}{\gamma}+\left(w\inv{\lambda}+(1-w)\inv{\gamma}\right)\inv{\gamma}\dfrac{(\mu-\lambda)(\gamma+\mu)}{\mu(\gamma+\mu-\lambda)}\dfrac{\gamma-\lambda}{\lambda}.
	\end{align*}
	With a little algebra, this expression can be simplified to that of Lemma \ref{lemma::ET1Y2}$\qedsymbol$
\end{myProof}
\section{Proof of Proposition \ref{prop::zw_aaoi}}\label{app::appendix3}
\begin{myProof}
	\begin{figure}[htpb]\captionsetup[subfigure]{font=footnotesize}
		\centering
		\subfloat[]{\includegraphics[width=0.48\textwidth]{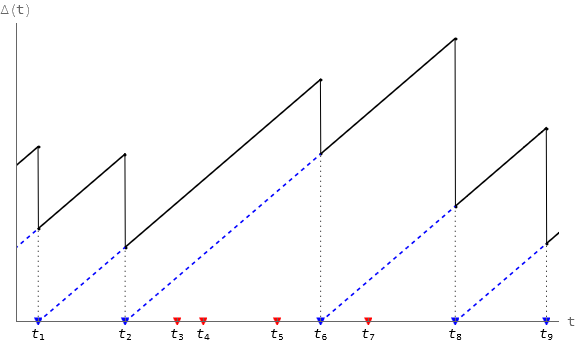}}\qquad
		\subfloat[]{\includegraphics[width=0.48\textwidth]{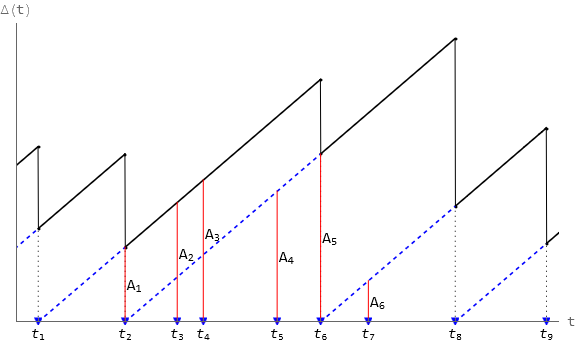}}
		\begin{tcolorbox}[colback=white,colframe=white,top=0mm,left=0mm,right=0mm]
			\footnotesize{Fig. \ref{fig::zw_model}.\quad Error-prone, zero-wait model as a system with aged updates. a) Error-prone, zero-wait standard model. Blue triangles are successful deliveries and generation of a fresh packet, and red triangles ($t_3$ and $t_5$) mark failed deliveries and subsequent retransmission of the same packet. b) Equivalent zero-wait, error-free model with aged updates. Failed transmissions in the standard model can be thought of as inducing an independent process of initial ages $A_n$.}
		\end{tcolorbox}
	\end{figure}
	
	\autoref{fig::zw_model} provides visual support for the demonstration. First, we note that the number of retransmissions for a given packet follows a geometric distribution with probability of success $(1-\alpha)$, $$p_k\eqdef\Pb{\text{k retransmissions}}=\alpha^k(1-\alpha),\quad k=0,1,2,\ldots$$ Consequently, the \emph{effective} service time in the zero-wait, error-prone model depicted in \autoref{fig::zw_model}a) is a random sum of $\expo{\mu}$ times, with the number of terms geometric of parameter $\alpha$. This leads to exponential effective service times with rate $\mu(1-\alpha)$.\\ Moreover, the probability that the $n$-th packet is fresh equals the probability that the previous delivery was successful; therefore $$\Pb{n\text{-th packet fresh}}=\Pb{(n-1)\text{-st delivery successful}}=(1-\alpha).$$
	From \autoref{fig::zw_model}a), we observe that the zero-wait error-prone model produces four types of packets, classified by whether they are fresh or retransmitted, and whether their delivery is successful or not. The initial age of each packet in the corresponding error-free model depends on its type in the original model. Below, we describe each type in the original model and deduce the corresponding initial age in the equivalent model.
	\begin{enumerate}
		\item[i)] A packet is fresh and successfully delivered with probability $(1-\alpha)^2$. These packets have zero initial age in the equivalent error-free model, like packets 1 and 8 in \autoref{fig::zw_model}a). Hence, $A_1=A_8=0$.
		\item[ii)] Packets 2 and 6 in \autoref{fig::zw_model}a) are fresh but have not been delivered, an event that occurs with a probability $\alpha(1-\alpha)$. In the error-free model, this is equivalent to a packet whose delivery does not change the age process, and therefore behaves as if it arrives with the same initial age as the monitor currently has. This age coincides with the system time of the immediately preceding packet, which in this model is also the effective service time. Hence, $A_2\sim A_6\sim\expo{\mu(1-\alpha)}$.
		\item[iii)] Packets 5 and 7 in \autoref{fig::zw_model}a) are retransmissions that are correctly delivered, occurring with a probability $\alpha(1-\alpha)$. In the error-free model shown in \autoref{fig::zw_model}b), this is equivalent to a packet arriving with an initial age equal to the time accumulated in the system by the original update. This is a geometric sum of $\expo{\mu}$ times where the probabilities are given by $\Pb{k\text{ retransmissions}}/\Pb{\text{retransmissions}}=p_k/(1-p_0)$. This yields $\expo{\mu(1-\alpha)}$ times for $A_5$ and $A_7$.
		\item[iv)] Forwarded packets that are not delivered, like packets 3 and 4 in \autoref{fig::zw_model}a), occur with a probability $\alpha^2$. The delivery of these packets does not change the age function, and thus, in the error-free model, it is equivalent to delivering a packet that enters the system with the same age as the monitor currently has. This age is the sum of the system time of the last successfully delivered packet --$\expo{\mu(1-\alpha)}$, as shown in part ii)-- and the time accumulated in the system by the original packet --also $\expo{\mu(1-\alpha)}$, as shown in part iii). Hence, $A_3\sim A_4\sim\Gamma\left(2,\mu(1-\alpha)\right)$. 
	\end{enumerate}
	Therefore,
	\begin{equation*}
		A_n=\begin{cases}0&\text{with probability }(1-\alpha)^2\\
			\expo{\mu(1-\alpha)}&\text{with probability }2\alpha(1-\alpha)\\
			\Gamma\left(2,\mu(1-\alpha)\right)&\text{with probability }\alpha^2\end{cases},
	\end{equation*}
	and $$\E{A_n}=\dfrac{2\alpha(1-\alpha)}{\mu(1-\alpha)}+\dfrac{2\alpha^2}{\mu(1-\alpha)}=\dfrac{2\alpha}{\mu(1-\alpha)}.$$
	Furthermore, the initial age process is independent of any other variable in the model because it consists of service times in the zero-wait, error-prone model, which are assumed to be independent. Thus, applying \autoref{thm::aaoi_au} to the equivalent error-free model results in $\Delta_{zw}^A=\Delta_{zw}^0+\E{A_n}$, where $\Delta_{zw}=2/\mu$ is the AAoI in a zero-wait, error free model with $\expo{\mu}$ service times. This yields $$\Delta_{zw}^A=\dfrac{2}{\mu}+\dfrac{2\alpha}{\mu(1-\alpha)}=\dfrac{2}{\mu(1-\alpha)}\qedsymbol$$
\end{myProof}
\section{Proof of Proposition \ref{prop::aaoi_hem1}}\label{app::appendix3}
\begin{myProof}[Proposition \ref{prop::aaoi_hem1}]
	This queue is a particular case of the G/M/1/$\infty$ queue, which has been extensively studied \cite{kleinrock:qst1,ross:ipm}. There are two main consequences of not having Poisson Process-type inputs. First, the PASTA (Poisson Arrivals See Times Averages) property is no longer available: the probability that an arrival sees the system in state $q$ is not the same as the stationary probability of the system being in state $q$. Second, since the inter-arrival times do not have the memoryless property, the residual inter-arrival times (which are also referred to here as idle times, since the server is idle during this period) and the real inter-arrival times do not have the same distribution. Finding the probability distribution of the residual inter-arrival times is another matter of interest.

	Losing the PASTA property means that, to study this system, not only the number of packets is needed, but also the time elapsed since the last arrival. To overcome this problem, the system is observed at very specific instants. Let $L_k$ be the number of packets in the system \emph{just before the $k$th arrival}. Then the stationary number of jobs in the system before arrival is $L=\lim_{k\to\infty}L_k$ and, by \cite{kleinrock:qst1}, Equation (6.27),	

	\begin{equation}\label{eq::hem1_Ldist}
		\Pb{L=n}=(1-\sigma)\sigma^n,\quad n=0,1,2,\ldots
	\end{equation}
	where $\sigma$ is the unique root in $(0,1)$ of the following equation, that involves the Laplace-Stieltjes Transform of the inter-arrival times:

	\begin{equation*}
		\sigma=\tilde X(\mu-\mu\sigma).
	\end{equation*}

	For the case at hand, $\tilde X(s)=\dfrac{\lambda}{\lambda+s}\dfrac{\gamma}{\gamma+s}$ and

	\begin{subequations}\label{def::sigma}
		\begin{equation}\label{def::sigma_eq1}
			\sigma=\dfrac{\lambda}{\lambda+\mu-\mu\sigma}\dfrac{\gamma}{\gamma+\mu-\mu\sigma},
		\end{equation}
		or
		\begin{equation}\label{def::sigma_eq2}
			\sigma=\dfrac{(\lambda+\gamma+\mu)-\sqrt{(\lambda+\gamma+\mu)^2-4\lambda\gamma}}{2\mu}.
		\end{equation}
	\end{subequations}

	Arguing as in the M/M/1/$\infty$ queue,it can be concluded that the system times are exponentially distributed with rate ($\mu-\mu\sigma$), $T_n\sim\expo{\mu-\mu\sigma}$. Therefore, Lemma \ref{lemma::3} gives the probability

	\begin{equation}\label{eq::X<T}
		\Pb{X_n<T_{n-1}} = \dfrac{\lambda}{\lambda+\mu-\mu\sigma}\dfrac{\gamma}{\gamma+\mu-\mu\sigma},
	\end{equation}
	which, by Equation \eqref{def::sigma_eq1}, is equal to $\sigma$. This result can be better understood by realizing that the event $\left\{X_n>T_{n-1}\right\}$ is the same as the event $\left\{\text{the $n$th arrival finds an empty system}\right\}$, and that the probability of the latter is $1-\sigma$ by the geometrical distribution \eqref{eq::hem1_Ldist}.

	Using Lemma \ref{lemma::4}a), the density of the residual inter-arrival times, defined as $$X^R_n\eqdef X_n-T_{n-1}|X_n>T_{n-1},$$ is given by

	\begin{equation}\label{eq::pdf_XR_hem1}
		f_{X^R}(t) = w\lambda\ex^{-\lambda t} + (1-w)\gamma\ex^{-\gamma t},\quad w=\dfrac{\gamma(\gamma+\mu-\mu\sigma)}{(\gamma-\lambda)(\lambda+\gamma+\mu-\mu\sigma)}
	\end{equation}

	By the Lindley-type identity \eqref{eq::Yn_identity} for the inter-departure times and Equation \eqref{eq::X<T}, it can be concluded that $Y_n$ is a service time with probability $\sigma$, or the sum of $S_n$ and $X^R_n$ with probability $(1-\sigma)$. With a little algebra, it can be proved that the convolution of $f_S(t)$ and $f_{X^R}(t)$ yields

	\begin{equation*}
		f_{S+X^R}(t) = w\dfrac{\mu\lambda}{\mu-\lambda}\left(\ex^{-\lambda t}-\ex^{-\mu t}\right) + (1-w)\dfrac{\mu\gamma}{\mu-\gamma}\left(\ex^{-\gamma t}-\ex^{-\mu t}\right)
	\end{equation*}
	which is a linear combination of $\Hexp{\lambda,\mu}$ and $\Hexp{\gamma,\mu}$ with the same weights of \eqref{eq::pdf_XR_hem1}. Hence, the density of $Y_n$ results in

	\begin{subequations}
		\begin{align}
			f_Y(t) &= \sigma f_S(t) + (1-\sigma) f_{S+X^R}(t)\nonumber\\
			&= a_1\mu\ex^{-\mu t} + a_2\lambda\ex^{-\lambda t} + a_3\gamma\ex^{-\gamma t},\nonumber\\
			\intertext{with $a_2=(1-\sigma)\dfrac{w\mu}{\mu-\lambda},\ a_3=(1-\sigma)\dfrac{(1-w)\mu}{\mu-\gamma},\ a_1=1-a_2-a_3$. The first two moments are}
			\E{Y_n}  &= \dfrac{1}{\lambda} + \dfrac{1}{\gamma}\label{eq::EY_hem1}\\
			\E{Y_n^2} &= 2\left(\dfrac{1}{\gamma^2} + \dfrac{1}{\lambda^2} + \dfrac{1+\sigma}{\lambda\gamma}\right)=\E{X_n^2}+\dfrac{2\sigma}{\lambda\gamma}\label{eq::EY2_hem1}
		\end{align}
	\end{subequations}

	From Lemma \ref{lemma::4}c), the joint probability of $X^R_n$ and $T^<_{n-1}\eqdef T_{n-1}|X_n>T_{n-1}$ and the expected value of their product are

	\begin{align*}
		\Pb{X^R_n>x,T^<_{n-1}>t} &= w\ex^{-\lambda x-(\lambda+\mu)-\mu\sigma)t} + (1-w)\ex^{-\gamma x-(\gamma+\mu)-\mu\sigma)t}\\&\\
		\E{X^R_nT^<_{n-1}} &= \int_{x\geq0}\int_{t\geq0}\Pb{X^R_n>x,T^<_{n-1}>t} \\
		&= \dfrac{1}{\lambda(\lambda+\mu-\mu\sigma)} + \dfrac{1}{\gamma(\gamma+\mu-\mu\sigma)}
	\end{align*}

	Thus, the expected value of the product $Y_nT_{n-1}$ is given by

	\begin{align}\label{eq::EYT_hem1}
		\E{Y_nT_{n-1}} &= \E{S_n}\E{T_{n-1}} + (1-\sigma)\E{X^R_nT^<_{n-1}}\nonumber\\
		&= \dfrac{1}{\mu}\left(\dfrac{1}{\gamma}+\dfrac{1}{\lambda}-\dfrac{1}{\mu}\right) + \dfrac{\sigma}{\mu}\left(\dfrac{1}{\gamma}+\dfrac{1}{\lambda}-\dfrac{1}{\mu-\mu\sigma}\right).
	\end{align}

	Plugging equations \eqref{eq::EY_hem1}, \eqref{eq::EY2_hem1} and \eqref{eq::EYT_hem1} into the definition of the AAoI

	\begin{equation*}
		\Delta_{HE/M/1/\infty} = \dfrac{\E{Y_n^2}/2 + \E{Y_nT_{n-1}}}{\E{Y_n}}
	\end{equation*}
	yields Proposition \ref{prop::aaoi_hem1}$\qedsymbol$
\end{myProof}
\section{Proof of Theorem \ref{thm::rq_solution}}\label{app::appendix4}
The steady-state solutions for this system can be obtained following the same procedure as in \cite[Section 3.5.1]{gross:fqt}. Let $p_{i,n}$ be the probability that the system is in state $(i,n)$. From the flow diagram of Figure \ref{fig::retrial_diag}, we can derive the balance equations for this system:

\begin{subequations}
	\begin{align}
		\lambda p_{0,0} &= \mu p_{1,0}\label{eq::baleq00}\\
		(\lambda+\theta) p_{0,n} &= \mu p_{1,n},\qquad {n\geq1}\label{eq::baleq0n}\\
		(\lambda+\mu)p_{1,0} &= \phantom{\lambda p_{1,n-1} +}\lambda p_{0,0} + \theta p_{0,1}\label{eq::baleq10}\\
		(\lambda+\mu)p_{1,n} &= \lambda p_{1,n-1} + \lambda p_{0,n} + \theta p_{0,n+1}\label{eq::baleq1n},\quad n\geq1
	\end{align}
\end{subequations} 

Define the partial Probability Generating Functions (PGF) 
$$P_0(z)=\sum\limits_{n=0}^\infty p_{0,n}z^n,\qquad P_1(z)=\sum\limits_{n=0}^\infty p_{1,n}z^n,$$
and recall that $p_{i,n}$ is the coefficient of $z^n$ in $P_i(z)$ and  $P_0(1)+P_1(1)=1$. Multiply \ref{eq::baleq0n} by $z^n$, sum over $n\geq 1$, and add \ref{eq::baleq00}:

\begin{subequations}\label{eq::pgf_eq_system}
	\begin{equation}\label{eq::pgf_eq1}
		(\lambda+\theta)P_0(z) - \theta p_{0,0} = \mu P_1(z).
	\end{equation}
	Doing the same with equations \ref{eq::baleq1n} and \ref{eq::baleq10}:
	\begin{equation}\label{eq::pgf_eq2}
		(\lambda+\mu-\lambda z)P_1(z) = (\lambda+\theta/z)P_0(z) - \theta p_{0,0}/z.
	\end{equation}
	On the other hand, by induction it can be shown that $\lambda p_{1,n} = \theta p_{0,n+1},\ n\geq 0$, which gives another relationship between PGF's:
	\begin{equation}\label{eq::pgf_eq3}
		\lambda P_1(z) = \theta P_0(z)/z - \theta p_{0,0}/z.
	\end{equation}
\end{subequations}

Let $P_i(1)=\sum\limits_{n=0}^\infty p_{i,n}=p_i,\ i=0,1$. Evaluating \eqref{eq::pgf_eq1} (or \eqref{eq::pgf_eq2}) and \eqref{eq::pgf_eq3} in $z=1$ yields

\begin{subequations}
	\begin{align}
		(\lambda+\theta)p_0 - \theta p_{0,0} &= \mu p_1\label{eq::peq1}\\
		\theta p_0 - \theta p_{0,0} &= \lambda p_1\label{eq::peq2}.
	\end{align}
\end{subequations}

This system results in $\lambda p_0 = (\mu-\lambda)p_1$, which together with $p_0+p_1=1$ finally gives the solutions

\begin{equation}\label{eq::server_ss_probs}
	p_1=\lambda/\mu\eqdef\rho,\qquad p_0=1-\rho.
\end{equation}

Define the probability that an update comes from the orbit, provided that it is non-empty, as $\pi=\theta/(\lambda+\theta)$. Using \eqref{eq::peq2} and solving for $p_{0,0}$ gives,

\begin{equation*}
	p_{0,0} = p_0-\lambda p_1/\theta = 1 - \rho/\pi.
\end{equation*}
Note that $\pi>\rho$ must hold. This can be regarded as a stability condition for this system. Substituting \eqref{eq::pgf_eq3} in \eqref{eq::pgf_eq2}, and \eqref{eq::p00} in \eqref{eq::pgf_eq1} yields

\begin{align*}
	P_1(z) &= \dfrac{\rho}{1-\rho z}P_0(z)\\
	P_1(z) &= \dfrac{\rho}{1-\pi}P_0(z) + \dfrac{\rho(\pi-\rho)}{1-\pi}.
\end{align*}

Equating both expressions and solving for $P_0(z)$ yields

\begin{equation*}
	P_0(z) = (\pi-\rho)\dfrac{1-\rho z}{\pi-\rho z} = (\pi-\rho) + \dfrac{(\pi-\rho)(1-\pi)}{\pi-\rho z},
\end{equation*}
and solving for $P_1(z)$ gives 

\begin{equation*}
	P_1(z) = \dfrac{\rho(\pi-\rho)}{\pi-\rho z}.
\end{equation*}

Given that $\dfrac{1}{1-\rho z/\pi}=\sum\limits_{n=0}^\infty(\rho/\pi)^nz^n$ for $|z|<\pi/\rho$, the proof is complete $\qedsymbol$
\end{document}